\documentclass{aa}

\usepackage{epsfig}
\input{epsf}

\newcommand{\Mat}[1]{{\bf#1}}
\newcommand{\tp}{{\rm t}}

\newcommand{\average}[1]{\left\langle #1 \right\rangle}

\newcommand{\ve}{\varepsilon}

\newcommand{\dd}{{\mathrm d}}

\newcommand{\del}{\partial}

\newcommand{\tot}{{\rm tot}}
\newcommand{\rul}{\rule[-2mm]{0mm}{5.5mm}}
\newcommand{\rulup}{\rule[-1mm]{0mm}{4.5mm}}
\newcommand{\ruldown}{\rule[-2mm]{0mm}{3.5mm}}

\newcommand{\Omegam}{\Omega_{\rm m}}

\newcommand{\J}{{\rm J}}

\newcommand{\Mapsq}{M_{\rm ap}^2}

\newcommand{\mi}{\phantom{-}}

\begin{document}

\title{Principal Component Analysis of Weak Lensing Surveys}

\author{Dipak Munshi\inst{1,2} \and Martin Kilbinger\inst{3}}

\institute{Institute of Astronomy, Madingley Road,
  Cambridge, CB3 OHA, United Kingdom\\
  \and
  Astrophysics Group, Cavendish Laboratory, Madingley Road, 
  Cambridge CB3 OHA, United Kingdom\\
  \and
  Argelander-Institut f\"ur Astronomie%
  \thanks{Founded by merging
    of the Sternwarte, Radio\-astro\-nomisches Institut and Institut f\"ur
    Astrophysik und Extraterrestrische Forschung der Universit\"at
    Bonn}
  , Universit\"at Bonn,
  Auf dem H\"ugel 71, D-53121 Bonn, Germany
}

\offprints{munshi@ast.cam.ac.uk}

\date{Received / Accepted}

\abstract%
%
{%
}%
{%
  We study degeneracies between cosmological parameters and
  measurement errors from cosmic shear surveys. We simulate realistic
  survey topologies with non-uniform sky coverage, and quantify the
  effect of survey geometry, depth and noise from intrinsic galaxy
  ellipticities on the parameter errors. This analysis allows us to
  optimise the survey geometry.
}%
{%
  We carry out a principal component analysis of the Fisher
  information matrix to assess the accuracy with which linear
  combinations of parameters can be determined. Using the shear
  two-point correlation functions and the aperture mass dispersion,
  which can directly be measured from the shear maps, we study various
  degeneracy directions in a multi-dimensional parameter space spanned
  by $\Omegam$, $\Omega_\Lambda$, $\sigma_8$, the shape parameter
  $\Gamma$, the spectral index $n_{\rm s}$, along with parameters that
  specify the distribution of source galaxies.
}%
{%
  A principal component analysis is an effective tool to probe the
  extent and dimensionality of the error ellipsoid. If only three
  parameters are to be obtained from weak lensing data, a single
  principal component is dominant and contains all information about
  the main parameter degeneracies and their errors.  For four or more
  free parameters, the first two principal components dominate the
  parameter errors. The degeneracy directions are insensitive against
  variations in the noise or survey geometry. The variance of the
  dominant principal component of the Fisher matrix, however, scales
  with the noise. Further, it shows a minimum for survey strategies
  which have small cosmic variance and measure the shear correlation
  up to several degrees. This minimum is less pronounced if external
  priors are added, rendering the optimisation less effective. The
  minimisation of the Fisher error ellipsoid can lead to slightly
  different results than the principal component analysis.
}%
{%
}%

%
  \keywords{ Cosmology: theory -- gravitational lensing -- large-scale
    structure of Universe -- Methods: analytical, statistical,
    numerical }

\titlerunning{PCA of weak lensing surveys}

\maketitle

\section{Introduction}


Recent observations by the Wilkinson Microwave Anisotropy Probe (WMAP)
mission confirmed the standard cosmological model with a very high
degree of accuracy (Spergel et al.~2003). In particular, these
observations confirmed that the universe is spatially flat and
dominated by dark energy and dark matter. Regarding the initial power
spectrum of scalar perturbations, the predictions of
the simplest inflationary models were strengthened, i.e.\ the near scale-invariance,
adiabaticity and Gaussianity of the initial density perturbations.
However, certain outstanding issues remain to be solved, such as the
running of spectral index $\alpha_{\rm s}$ which can be
addressed in more detail with additional data from galaxy surveys such
as SDSS (York et al.\ 2000), 2dF (Colless et al.\ 2001) and the
Lyman-$\alpha$ forest (see Seljak et al.\ 2003 and references
therein).


Weak lensing surveys are expected to make important and complementary
contributions to high-precision measurements of cosmological parameters.
Contaldi et al.\ (2003) used the Red Cluster Sequence (RCS) to show that the
$\Omegam$-$\sigma_8$ degeneracy direction is nearly orthogonal to the one from
CMB measurements, making weak lensing particularly suitable for combined
analyses (van Waerbeke et al.\ 2002). Ishak et al. (2003) argued that a joint
CMB-cosmic shear survey provides an optimal data set for constraining the
amplitude and running of spectral index which helps to probe various
inflationary models. Tereno et al.\ (2004) studied cosmological forecasts for
joint CMB and weak lensing data. Clearly, the potential of weak lensing
surveys (Mellier 1999; Bartelmann \& Schneider 2001; R\'efr\'egier 2003; van
Waerbeke \& Mellier 2003; Schneider 2005) as a cosmological probe is now well
established (Contaldi et al. 2003; Hu \& Tegmark 1999). In the last few years
there have been many studies which have detected cosmic shear in random
patches of the sky (Brown et al. 2003; Bacon et al. 2003; Bacon, R\'efr\'egier
\& Ellis 2000; Hamana et al. 2003; H\"{a}mmerle et al. 2002; Hoekstra et al.
2002a; Hoekstra, Yee \& Gladders 2002a; Jarvis et al.  2002; Kaiser, Wilson \&
Luppino 2000; Maoli et al.  2001; R\'efr\'egier, Rhodes, \& Groth 2002;
Rhodes, R\'efr\'egier \& Groth 2001; van Waerbeke et al. 2000, 2001, 2002;
Wittman et al. 2000).  While early studies were primarily concerned with the
detection of the weak lensing signal, present generations of weak lensing
observations are putting constraints on cosmological parameters, in particular the
matter density parameter $\Omegam$ and the power spectrum normalisation
$\sigma_8$.


Inspired by the success of these surveys, there are many other ongoing,
planned and proposed weak lensing surveys which are currently in progress,
including the Deep Lens Survey (Wittman et al.~2002), the Canada-France-Hawaii
Telescope Legacy Survey (Hoekstra et al.~2005; Semboloni et al.~2005), the
Panoramic Survey Telescope and Rapid Response System, the Supernova
Acceleration Probe (Massey et al.~2003), the NOAO Deep Wide-Field Survey
(Groch et al.~2002) and the Large Synoptic Survey Telescope (Tyson et
al.~2002).  Future cosmic shear surveys will be able to probe much larger
scales in the linear regime which will provide more stringent bounds on
cosmological parameters such as the equation of sate of dark energy and its
time variations.

In a recent work (Kilbinger \& Schneider 2004, hereafter KS04), the impact of
the survey design on cosmological parameter constraints was analysed using a
likelihood and Fisher matrix analysis, extending previous studies based on the
assumption of uniform sky coverage (Schneider et al.~2002a). Earlier work in
this direction by Kaiser (1998) considered a singe $3^\circ \times 3^\circ$
-field and studied the effect of sparse sampling and intrinsic ellipticity
dispersion.  The motivation for the present work remains the same, although we
concentrate on the eigenvalues of the Fisher matrix.  Therefore, in contrast
to KS04 where the 1$\sigma$ errors on individual parameters have been used to
optimise the survey geometry, we consider all parameter combinations
corresponding to the eigenvectors of the Fisher matrix.  We study how noise
due to the intrinsic ellipticity dispersion of galaxies $\sigma_\epsilon$, the
number density of galaxies $n_{\rm gal}$, the survey depth and marginalisation affects the
determination of the parameter combinations for various survey strategies.


Cosmic shear is sensitive to a large number of cosmological
parameters. However, the dependency on these parameters is partially
degenerate (although these degeneracies can be broken
by the use of external data sets such as CMB, galaxy surveys and
Lyman-$\alpha$ surveys).
A principal component analysis (PCA) can be used as an efficient tool to
identify the degeneracy directions and linear combinations of
cosmological parameters, rank-ordered according to the accuracy with which they can be
determined from a given survey set-up. Indeed, in recent years there has been
a renewed interest in applying principal component analysis techniques to
various cosmological data sets, a technique pioneered by Efstathiou \& Bond
(1999).  This method can reveal the detailed statistical structure of
cosmological parameter space which is lacking in an one-dimensional confidence
level presentation.  Efstathiou (2002) studied PCA in the context of the
tensor degeneracy in CMB. For a recent work see Rocha et al.\ (2004), where
the possibility of measurement of the fine-structure constant $\alpha$ has
been explored in the context of CMB data with analysis based on Fisher matrix
and PCA.  Hu \& Keeton (2002) applied this technique to map the density
distribution along the radial direction from weak lensing surveys.  Jarvis \&
Jain (2004) used PCA to correct for the point spread function (PSF) variation
in weak lensing surveys.  In the context of SN Ia observations to constrain the
dark energy equation of state, Huterer \& Starkman (2003) and Huterer \&
Cooray (2004) employed PCA and its variants (see Wang \& Tegmark 2005;
Crittenden \& Pogosian 2005 for more recent results). Tegmark et al.~(1998)
used PCA for decorrelating the power spectrum of galaxies. This idea was
initially proposed by Hamilton (1997) and further discussed in the context of
galaxy surveys by Hamilton \& Tegmark (2000).


This paper is organised as follows: in Sect.~\ref{sec:notation}, a
very brief overview of our notations is provided; in particular, we
introduce how the covariance matrix and the Fisher matrix is
constructed for a given estimator and a given survey strategy. This
section also outlines the basics of principal components analysis.  In
the next section (Sect.~\ref{sec:numbers}) we provide the details of
survey geometries and the numerical results of the PCA considering
three survey set-ups. For a small number of cosmological parameters,
($\le 4$) we consider various survey strategies and try to optimise
those. For a larger set of parameters, we consider a ten times larger shear survey.
Effects of various noise sources on the principal components are
investigated. Section~\ref{sec:discussion} is left for discussions and
future prospects.

\section{Notation and formalism}
\label{sec:notation}

\subsection{Second-order shear statistics}
\label{sec:shear_est}
\label{sec:cov_est}

In our numerical studies presented here, we use the two-point
correlation functions of shear $\xi_{\pm}$ and the aperture mass
dispersion $\langle M_{\rm ap}^2 \rangle$ to predict constraints on
cosmological parameters. Both these statistics depend linearly on the
convergence power spectrum $P_\kappa$ (Kaiser 1992; Kaiser 1995; Schneider 1996;
Schneider et al.~1998)
\begin{eqnarray}
  \xi_{\pm}(\theta) & = & \frac{1}{2 \pi} \int\limits_0^\infty \dd \ell
  \, \ell \, P_\kappa(\ell) \J_{0,4}(\ell \theta); \nonumber \\
  \average{ \Mapsq(\theta)} & =  &\frac{1}{2\pi} \int\limits_0^\infty \dd
  \ell \, \ell \, P_\kappa(\ell) \left( \frac{24 \, \J_4(\ell \theta)}{(
      \ell \theta )^2} \right)^2,
  \label{xi-Map-def}
\end{eqnarray}
where $\J_\nu$ is the first-kind Bessel function of order $\nu$.

Estimators of these statistics and their covariances are defined in
Schneider et al.~(2002a). We use the Monte-Carlo-like method from KS04
to integrate the analytical expressions of the covariances, which are
exact in case of a Gaussian shear field. Our result is expected to
underestimate the covariance due to non-Gaussian
contributions on scales between $\sim$ 1 and 10 arc minutes.

\subsection{Fiducial cosmological model}
\label{sec:cosmo}

We calculate the convergence power spectrum and the shear estimators using the
non-linear fitting formulae of Peacock \& Dodds (1996). Our cosmological
model has seven free parameters: These are the five cosmological parameters
$\Omega_{\rm m}$, $\Omega_\Lambda$, the power spectrum normalisation
$\sigma_8$, the spectral index of the initial scalar fluctuations $n_{\rm s}$
and $\Gamma$, which determines the shape of the power spectrum. The fiducial
model is assumed to be a flat $\Lambda$CDM cosmology with $\Omegam = 0.3$,
$\sigma_8 = 1$, $\Gamma = 0.21$ and $n_{\rm s} = 1$. The two parameters $z_0$
and $\beta$ characterise the redshift
distribution of background galaxies (Brainerd et al.~1996),
\begin{equation}
p(z) {\rm d} z = { \beta \over z_0 \Gamma(3/\beta) } \left ({ z \over z_0}
\right)^2 e^{-(z/z_0)^{\beta}} {\rm d} z,
\label{pzdz}
\end{equation}
with fiducial values $z_0 = 1$ and $\beta = 1.5$.

\subsection{Principal components analysis of the Fisher matrix}
\label{sec:decor}

We use the expression for the Fisher matrix (see Tegmark, Taylor \& Heavens
1997 for a review) in the case of Gaussian errors and parameter-independent
covariance,
\begin{equation}
F_{ij} = \sum\limits_{ij} \left( \frac{ \del x_k }{ \del \Theta_i } \right)
(\Mat C^{-1})_{kl} \left( \frac{ \del x_l }{ \del \Theta_j } \right),
\end{equation}
where $x_k$ is either $\xi_+(\theta_k)$, $\xi_-(\theta_k)$, $\langle
M_{\rm ap}^2(\theta_k) \rangle$ or an entry of the combined
correlation function $\xi_{\rm tot} = (\xi_+, \xi_-)$. $\Mat C$
denotes the covariance matrix of the estimator of the corresponding
shear statistics, $\vec \Theta = (\Theta_1, \ldots \Theta_{n})$ is the
vector of cosmological parameters. The inverse of the Fisher matrix is
the covariance of the parameter vector at the point of maximum
likelihood,
\begin{equation}
\Mat F^{-1} = \langle \Delta \vec \Theta \Delta \vec \Theta^\tp \rangle
= \langle \vec \Theta \vec \Theta^\tp \rangle - \langle
\vec \Theta \rangle \langle \vec \Theta^\tp \rangle.
\end{equation}
The standard deviation of the $i^{\rm th}$ parameter obtained from the
Fisher matrix, $\Delta \Theta_i = (\langle\Theta_i^2\rangle -
  \langle \Theta_i \rangle^2)^{1/2} = [(\Mat F^{-1})_{ii}]^{1/2}$, is
called the minimum variance bound (MVB). According to the Cram\'er-Rao
inequality, the variance of any unbiased estimator is always larger or
equal to the MVB.

Any real matrix ${\Mat W}$ is called a {\it decorrelation matrix} if
it satisfies
\begin{equation}
  {\Mat F} = {\Mat W}^\tp \Mat \Lambda \Mat W,
  \label{Fdec}
\end{equation}
where $\Mat \Lambda$ is a diagonal matrix (Hamilton \& Tegmark 2000).
The quantities ${ \vec \Phi } = \Mat W { \vec \Theta }$
are uncorrelated because their covariance matrix is diagonal,
\begin{equation}
  \langle
  {\Delta \vec \Phi} {\Delta \vec \Phi}^\tp \rangle = \Mat W \langle
  \Delta \vec \Theta \Delta \vec \Theta^\tp \rangle \Mat W^\tp = \Mat
  \Lambda^{-1}.
  \label{phi}
\end{equation}
By multiplying $\Mat W$ with the square root of the diagonal matrix
$\Mat \Lambda$, the quantities $\vec \Phi$ can be scaled to unit variance
without loss of generality. In this case, (\ref{Fdec}) is written as
\begin{equation}
 {\Mat F} = \tilde{\Mat W}^\tp \tilde{\Mat W},
  \label{Fdectilde}
\end{equation}
where $\tilde {\Mat W} = {\Mat \Lambda}^{1/2} \, \Mat W$.  Note that
the choice of $\tilde {\Mat W}$ is not unique. If some matrix
$\tilde{\Mat W}$ satisfies (\ref{Fdectilde}), the same is true for any
orthogonal rotation $\Mat O \tilde{\Mat W}$ with $\Mat O \in {\rm
  SO}(n)$ and therefore, there are infinitely many decorrelation
matrices satisfying (\ref{Fdectilde}).

If $\Mat W$ is an orthogonal matrix, its rows are the eigenvectors
$\vec p_i$ of $\Mat F$ and $\Mat \Lambda = {\rm diag}(\lambda_i)$ is
the diagonal matrix of the corresponding eigenvalues. In that case,
(\ref{Fdec}) is called a principal component decomposition. We assume
the eigenvalues to be in descending order.

The eigenvectors or principal components of $\Mat F$ determine the
principal axes of the $n$-dimensional error ellipsoid in parameter
space. The eigenvectors represent orthogonal linear combinations of
the physical (cosmological) parameters that can be determined
independently from the data. The more these vectors are aligned with
the parameter axes, the less are the degeneracies between those
parameters. The accuracy with which these linear parameter
combinations can be determined is quantified by the variance
$\sigma_i \equiv\sigma(\vec p_i)  = \Delta \Phi_i = \Lambda_{ii}^{-1/2} =
\lambda_i^{-1/2}$. Thus, a principal component decomposition of the
Fisher matrix gives us information about which (linear) parameter
combinations can be determined with what accuracy from a given data
set. Since the eigenvalues are in descending order, the first
eigenvector $\vec p_1$ having the smallest variance corresponds to the best
constrained parameter combination. The last eigenvector $\vec p_n$ is
the direction with the largest uncertainty.

From (\ref{phi}), one can calculate the MVB from the eigenvectors and
eigenvalues of $\Mat F$,
\begin{equation}
\Delta \Theta_j =
  \left( \sum_{i=1}^n W_{ji}^2 \, \lambda_i^{-1} \right)^{1/2} .
  \label{sum}
\end{equation}

\section{Numerical results}
\label{sec:numbers}

\subsection{Survey strategies}
\label{sec:survey}

We simulate shear surveys consisting of $P$ circular, uncorrelated
patches of radius $R$ on the sky, each in which $N$ individual fields
of view of size $13^\prime \times 13^\prime$ are distributed randomly
but non-overlapping. The total number of fields of view is $n = P N =
300$, corresponding to a total survey area of $A = 14.1$ square
degree. Different surveys with $N=$ 10, 20, 30, 50, and 60 are
considered, corresponding to geometries with $P=$ 30, 15, 10, 6 and 5
patches, respectively (see KS04).

We denote these survey strategies with $(N,R)$, e.g. $(50,100^\prime)$
corresponds to a survey with $N=50, R=100^\prime$ and $P=6$. Further, we
consider a survey consisting of 300 uncorrelated lines of sight \`a $13^\prime
\times 13^\prime$, which are randomly distributed on the sky. This survey,
denoted by $300\cdot13^{\prime \, 2}$, has smaller cosmic variance than any of
the patch strategies, but does not sample intermediate and large angular scales.

If not indicated otherwise, the number density of source galaxies is
$n_{\rm gal} = 30 \, {\rm arcmin}^{-2}$. This number density of
high-redshift galaxies which are usable for weak lensing shape
measurements can be achieved with high-quality ground-based imaging
data on a 4 m-class telescope.  The source galaxy ellipticity
dispersion is $\sigma_\ve = 0.3$, if not stated otherwise.  For
comparison, these quantities are varied to $n_{\rm gal} =20$ and $40 \,
{\rm arcmin}^{-2}$, and $\sigma_\ve = 0.2, 0.4$, respectively, to
study the effect of noise sources on the principal components.

\subsection{Eigenvalues of the Fisher matrix}
\label{sec:eigenval}

\begin{figure}
  \resizebox{1.0\hsize}{!}{
    \includegraphics[bb=18 144 592 730] {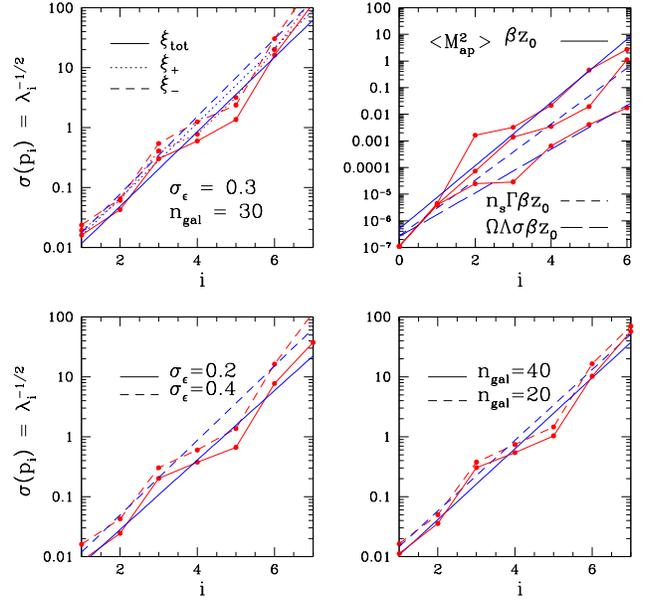}
  }
  \caption{The variance $\sigma(\vec p_i) = \lambda_i^{- 1/2}$ associated with 
    the principal components of the 7$\times$7 Fisher matrix. The
    straight lines are fits to the data points.
    \emph{Lower left panel:} $\sigma(\vec p_i)$ as a function of the
    intrinsic ellipticity dispersion of galaxies $\sigma_\varepsilon=0.2$
    and 0.4.
    \emph{Lower right:} the variation of $\sigma(\vec p_i)$ with the number
    density of galaxies $n_{\rm gal}=20$ and 40.
\emph{Upper left:} the variation of $\sigma(p_n)$ for three
        different estimators, $\xi_{+}$, $\xi_{-}$ and $\xi_{tot}$.
\emph{Upper right:} the variation of $\sigma(\vec p_i)$ for three
different priors, see Sect.~\ref{sec:priors}. The survey strategy is
$(30,100^\prime)$ in all cases.
}
  \label{fig:lambda_scale}
\end{figure}

We consider the Fisher matrix $\Mat F$ corresponding to all seven
cosmological and redshift parameters. In this section, we demonstrate
the influence of survey characteristics (other than the geometry) on
the eigenvalues $\lambda_i$ of $\Mat F$. The variance of the linear
combination of parameters given by the $i^{\rm th}$ eigenvector $\vec
p_i$ of $\Mat F$ is $\sigma(\vec p_i) = \lambda_i^{-1/2}$, as defined
in Sect.~\ref{sec:decor}. In Fig.~\ref{fig:lambda_scale} we show the
effect of the number of background galaxies, $n_{\rm gal}$, and the
intrinsic ellipticity dispersion, $\sigma_\ve$. In both cases, the noise
variation causes a scaling of the variance. For the intrinsic
ellipticity dispersion, the scaling factor increases with $i$. The
$i=1$ variance scales linearly with $\sigma_\ve$, whereas the mean
dependence (averaged over all 7 eigenvalues) is quadratic in
$\sigma_\ve$. In the case of $n_{\rm gal}$, however, the variance
$\sigma_i$ of all eigenvectors is scaled by a constant factor which is
inversely proportional to $n_{\rm gal}$.  The variance of the
eigenvectors are steeper functions of the noise characteristics than
the MVB (Kilbinger \& Munshi 2005). Note however, that in this
previous study the MVB was calculated for each parameter individually,
without taking parameter correlations into account.

\begin{figure}
\protect\centerline{
\epsfysize = 1.9truein
\epsfbox[30 430 590 719]
{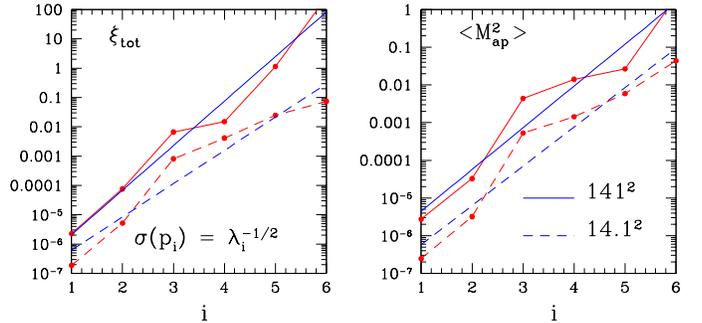}}
\caption{The variation of $\sigma(\vec p_i)$ 
  as a function of the eigenvector number $i$, for $\xi_{\rm tot}$
  (left panel) and $\langle M^2_{\rm ap} \rangle$ (right panel). The
  redshift parameter $\beta$ is fixed, and a Gaussian prior with
  variance $s(\sigma_8) = 0.1$ is added to regularise the Fisher
  matrix.  Two surveys with size $14.1$ (solid lines) and 141 square
  degree (dashed), respectively, are displayed.}
\label{fig:fsky}
\end{figure}

If boundary effects due to the finite field of the survey are
neglected, the covariance is anti-proportional to the observed survey
area.  Consequently, the variance $\sigma_i$ scales as $f_{\rm
  sky}^{-1/2}$, where $f_{\rm sky}$ is the fractional 
sky coverage of
the survey.  As an example, we compare the $300\cdot13^{\prime \, 2}$
survey with a survey consisting of 50 patches with $N=60$ and
$R=140^\prime$ (corresponding to ten $(60,140^\prime)$ surveys). We
found a good agreement on the expected scaling of $\sigma_i$ as a
function of the survey area, although for the combinations which are
worst constrained by the data, the dependence on $f_{\rm sky}$ seems
to be steeper (see Fig.~\ref{fig:fsky}).

\subsection{Eigenvectors of the Fisher matrix}
\label{sec:eigenvec}

In Tables \ref{table:map-fOGs}--\ref{table:xi-OGsL}, the eigenvectors
of the Fisher matrix corresponding to various combinations of parameters
are shown, for the 2PCF and the aperture mass dispersion.

\begin{table}[th]
\begin{center}
\caption{Eigenvectors of the Fisher matrix corresponding to
  ($\Omegam$, $\Gamma$, $\sigma_8$) and a flat Universe as prior, for
  $\langle M_{\rm ap}^2 \rangle$ using the survey strategy
  $(50,100^\prime)$. $\lambda_i^{-1/2} = \sigma_i$ is the variance of
  the $i^{\rm th}$ eigenvector, $\Theta_j$ the MVB for the $j^{\rm
  th}$ cosmological parameter. }
\label{table:map-fOGs}
\begin{tabular} {@{}lc|ccc|c}
\hline
\hline
&   &  $\vec p_1$   &  $\vec p_2$  &  $\vec p_3$  & $\triangle \Theta_j$  \\
\hline
& $\Omegam$ &    $0.649$ &  $-0.533$     &  $-0.542$ &            $0.165$\\
& $\Gamma$ &    $0.563$ &  $\mi 0.816$  &  $-0.128$ &            $0.042$ \\
& $\sigma_8$ &  $0.511$ &  $-0.222$     &  $\mi 0.830$  &            $0.252$ \\
\hline
& \rul $\lambda_i^{-1/2}$ & $0.004$ & $\mi 0.020$ & $
\mi 0.300$ &  \\
\hline
\hline
\end{tabular}
\end{center}
\end{table}

\begin{table}[th]
\begin{center}
\caption{Eigenvectors of the Fisher matrix corresponding to
  ($\Omegam$, $\Gamma$, $\sigma_8$, $n_{\rm s}$) and the prior
  $\Omega_\Lambda = 0.7$, for $\langle M_{\rm ap}^2 \rangle$ using the
  survey strategy $(50,100^\prime)$.}
\label{table:map-OGsn}
\begin{tabular} {@{}lc|cccc|c}
\hline
\hline
&  &  $ \vec p_1$   &  $\vec p_2$  &  $\vec p_3$ & $\vec p_4$ &
$\Delta \Theta_j$  \\
\hline
 & $\Omegam$ & 0.563 &  $-0.671$  &    $\mi 0.322$  &    $-0.358$ & 0.254 \\
 & $\Gamma$ & $0.575$  & $\mi 0.622$  &   $-0.235$  &   $-0.474$ & 0.333 \\
 & $\sigma_8$ & $0.522$  &   $-0.161$  &    $\mi 0.501$  &    $\mi 0.670$ & 0.473  \\
 & $n_{\rm s}$ &  0.281   &  $\mi 0.368$   &   $\mi 0.767$  &    $\mi 0.443$ & 0.329  \\
\hline
& \rul $\lambda_i^{-1/2}$ & $0.004$ & $\mi 0.015$ & $\mi 0.147$ & $\mi
0.698$ &   \\
\hline
\hline
\end{tabular}
\end{center}
\end{table}

\begin{table}
\begin{center}
  \caption{Eigenvectors of the Fisher matrix corresponding to
    ($\Omegam$, $\Gamma$, $\sigma_8$, $\Omega_{\Lambda}$) for $\langle
    M_{\rm ap}^2 \rangle$  using the survey strategy
    $(50,100^\prime)$.}
\label{table:map-OGsL}
\begin{tabular} {@{}lc|cccc|c}
\hline
\hline
& &  $\vec p_1$   &  $\vec p_2$  &  $\vec p_3$ & $\vec p_4$ & $\triangle \Theta_j$  \\
 \hline
& $\Omegam$            & $\mi 0.587$  &    $\mi 0.643$   &  $\mi 0.352$  &  $ -0.341$ & $0.169$ \\
& $\Gamma$            &   $\mi 0.591$  &   $ -0.694$  &   $\mi 0.392$ &   $\mi 0.114$& $ 0.112$  \\
& $\sigma_8$           &   $\mi 0.542 $  &  $\mi 0.117$  &   $ -0.727$ & $\mi 0.404$ & $0.252$ \\
&  $\Omega_{\Lambda}$ &    $-0.102$ &   $\mi 0.299$  &  $\mi 0.439$ &  $\mi 0.840$ & $0.368$
\\ \hline
& \rul $\lambda_i^{-1/2}$ & $\mi 0.004$ & $\mi 0.015$ & $\mi 0.258$ & $\mi 0.417$ &   \\
\hline
\hline
\end{tabular}
\end{center}
\end{table}

\begin{table}
\begin{center}
\caption{Eigenvectors of the Fisher matrix corresponding to
  ($\Omegam$, $\Gamma$, $\sigma_8$) and a flat Universe as prior, for
  $\xi_{\tot}$ ($\xi_+, \xi_-$ in brackets) using the survey strategy
  $(50,100^\prime)$.}
\label{table:xi-fOGs}
\begin{tabular} {@{}l|ccc|c}
\hline
\hline
 &  $\vec p_1$   &  $\vec p_2$  &  $\vec p_3$ &  $\triangle \Theta_j$ \\
\hline
$\Omegam^{}$  &  0.702        &    $-0.474$         &    $-0.530$        & 0.144\\
             & $(0.731)^+$   &    $(-0.435)^+$   &   $(-0.524)^+$      & $(0.164)^+$  \\
             & $(0.665)^-$   &   $(-0.515)^-$    &    $(-0.540)^-$     & $(0.213)^-$\\
\hline
$\Gamma$      &  $0.473$    &    $\mi 0.868$       &   $-0.150$           & 0.043\\
              &$(0.412)^+$  &   $\mi (0.895)^+$    &  $(-0.167)^+$       & $(0.056)^+$ \\
              & $(0.536)^-$ &    $\mi (0.833)^-$   &   $(-0.134)^-$      & $(0.057)^-$  \\
\hline
$\sigma_8$    &  $ 0.531$     &    $-0.145 $      &    $\mi 0.834$             & 0.226\\
              & $(0.542)^+$ &  $(-0.093)^+$   &   $\mi (0.834)^+$       & $(0.261)^+$ \\
              &$(0.519)^-$  &  $(-0.200)^-$   &   $\mi (0.830)^-$       &  $(0.327)^-$
\\ \hline
 \rulup $\lambda_i^{-1/2}$        &   0.004       &    $\mi 0.017$  & $\mi 0.271$ & \\
 $(\lambda_i^{-1/2})^+$   &  $(0.005)^+$   &   $\mi(0.020)^+$ & $\mi (0.313)^+$ & \\
 \ruldown $(\lambda_i^{-1/2})^-$   &  $(0.006)^-$   &   $\mi (0.027)^-$ & $\mi (0.394)^-$ & \\
\hline
\hline
\end{tabular}
\end{center}
\end{table}

\begin{table*}
\begin{center}
\caption{Eigenvectors of the Fisher matrix corresponding to
  ($\Omegam$, $\Gamma$, $\sigma_8$, $n_{\rm s}$) and the prior
  $\Omega_\Lambda = 0.7$, for $\xi_{\rm tot}$ ($\xi_+, \xi_-$ in
  brackets) using the survey strategy $(50,100^\prime)$.}
\label{table:xi-OGsn}
\begin{tabular} {@{}l|cccc|c}
\hline
\hline
        &  $\vec p_1$   &  $\vec p_2$  &  $\vec p_3$  & $\vec p_4$ &  $\triangle \Theta_j$  \\
\hline $\Omegam$ &  0.657   &   $-0.588$         & $\mi 0.282$   &   $-0.376$   & 0.154 \\
            &$( 0.708)^+$   &    $(-0.533)^+$    & $\mi (0.320)^+$  &    $(-0.332)^+$ & $(0.158)^+$  \\
            &$( 0.591)^-$   &   $(-0.647)^-$     & $\mi(0.329)^-$   &   $(-0.349)^-$ & $(0.290)^-$\\
\hline
$\Gamma$ &    $0.474$       &    $ \mi 0.719 $   & $-0.217$      &  $-0.459$ & 0.182\\
              & $(0.406)^+$ &    $\mi (0.768)^+$ & $(-0.116)^+$  & $(-0.480)^+$ & $(0.198)^+$\\
              & $(0.547)^-$ &   $\mi(0.654)^-$      & $(-0.205)^-$  & $(-0.480)^-$  & $(0.389)^-$\\
\hline
$\sigma_8$   &     $0.542$ &   $-0.062$         & $-0.474$        & $\mi 0.690$ & 0.279\\
             & $(0.548)^+$  &  $\mi (0.005)^+$   & $(-0.572)^+$    & $\mi (0.611)^+$ & $(0.289)^+$ \\
             &$(0.533)^-$   &  $(-0.129)^-$      & $(-0.522)^-$    &   $\mi(0.653)^-$ & $(0.538)^-$\\
\hline
$n_{\rm s}$  &  $0.217$  &   $\mi 0.363$  &   $0.805$ & $\mi 0.415$ & 0.209\\
             &$(0.179)^+$ & $\mi (0.352)^+$ &  $\mi (0.746)^+$ & $\mi (0.535)^+$ & $(0.290)^+$\\
             &$(0.257)^-$ &   $\mi(0.367)^-$ &   $\mi(0.760)^-$  & $\mi(0.470)^-$ & $(0.414)^-$\\
\hline
 \rulup $\lambda_i^{-1/2}$   &   0.004 &    $\mi 0.013$  & $\mi 0.165$ & $\mi 0.389$ \\
 $(\lambda_i^{-1/2})^+$   &  $(0.005)^+$   & $\mi(0.016)^+$ & $\mi(0.255)^+$ & $\mi(0.408)^+$ \\
 \ruldown $(\lambda_i^{-1/2})^-$   &  $(0.006)^-$   & $\mi(0.020)^-$ & $\mi(0.221)^-$ & $\mi(0.804)^{-}$\\
\hline
\hline
\end{tabular}
\end{center}
\end{table*}

\begin{table*}
\begin{center}
\caption{Eigenvectors of the Fisher matrix corresponding to
  ($\Omegam$, $\Gamma$, $\sigma_8$, $\Omega_\Lambda$), for $\xi_{\rm
    tot}$ ($\xi_+, \xi_-$ in brackets) using the survey strategy
  $(50,100^\prime)$.}
\label{table:xi-OGsL}
\begin{tabular} {@{}l|cccc|c}
\hline
\hline
 &  $\vec p_1$   &  $\vec p_2$  &  $\vec p_3$  & $\vec p_4$  & $\triangle \Theta_j$  \\
\hline
$\Omegam$ &  $\mi0.676$  &   $\mi 0.556$             &   $\mi 0.082$  &  $-0.475$ & 0.147 \\
         & $\mi (0.722)^+ $    &  $\mi (0.501)^+$   &   $\mi (0.030)^+$  & $(-0.475)^+$ & $(0.172)^+$ \\
         & $\mi (0.613)^- $    &  $(0.619)^-$   &   $\mi(0.348)^-$  &  $(-0.343)^-$ & $(0.213)^-$ \\
\hline
$\Gamma$  &    $\mi0.479$      &   $-0.773$       &  $\mi 0.385$        & $-0.156$     & 0.094 \\
          &  $\mi (0.408)^+ $  &     $(-0.808)^+$  &  $\mi (0.371)^+$    &   $(-0.207)^+$ & $(0.112)^+$ \\
          &  $\mi (0.558)^-  $ &  $(-0.718)^-$    &   $\mi(0.400)^-$   &   $\mi(0.110)^-$  & $(0.156)^-$ \\
\hline
$\sigma_8$  &   $\mi0.555$      &  $\mi 0.022$      &    $-0.339$         & $\mi 0.759$      & 0.244 \\
            & $\mi (0.556)^+ $  &  $(-0.039)^+$  &    $(-0.260)^+$   &   $\mi (0.788)^+$ & $(0.290)^+$ \\
            & $\mi (0.549)^-  $  &  $\mi(0.088)^-$   &  $(-0.721)^-$    &  $\mi(0.441)^-$ & $(0.333)^-$ \\
\hline
$\Omega_\Lambda$  &  $-0.060$   &    $\mi0.303$       &   $\mi 0.854$ &    $\mi 0.417$      & 0.220\\
             & $ (-0.032)^+ $  &   $\mi(0.307)^+$  &   $\mi (0.891)^+$  & $\mi (0.332)^+$ & $(0.232)^+$ \\
             & $ (-0.092)^- $  &   $\mi(0.304)^-$  &    $\mi(0.445)^-$   &   $\mi(0.837)^-$ & $(0.448)^-$\\
\hline
\rulup ${\lambda}_i^{-1/2}$  &   $\mi0.005$   &   $\mi 0.014$      & $\mi 0.210$ &  $\mi 0.308$ & \\
$(\lambda_i^{-1/2})^+$    &  $\mi(0.005)^+$   &   $\mi(0.017)^+$     & $\mi (0.223)^+$  & $\mi(0.362)^+$ &  \\
 \ruldown $(\lambda_i^{-1/2})^-$   &  $\mi(0.006)^-$   &   $\mi(0.021)^-$  &    $\mi(0.364)^-$ & $\mi(0.499)^-$ &  \\
\hline
\hline
\end{tabular}
\end{center}
\end{table*}

The best determined eigenvector $\vec p_1$ is always orthogonal to the
$\Omegam$-$\sigma_8$ degeneracy direction. The variance $\sigma_n$ of
the worst constrained principal component dominates the uncertainty of
all eigenvectors. In the case of $(\Omegam, \sigma_8, \Gamma$),
$\sigma_3^2 = 1/\lambda_3$ constitutes more than $99\%$ of the total
uncertainty (see Tables \ref{table:map-fOGs} and \ref{table:xi-fOGs}),
and therefore dominates the error on all cosmological parameters. For
$\Omegam$ and $\sigma_8$, the MVBs can be approximated using this
principal component alone, by $\sigma_3 |p_{j3}| \approx \Delta
\Theta_j$ where $j$ denotes the $j^{\rm th}$ parameter. This
corresponds to considering only the last ($i=3$) term on the
right-hand side of eq.~(\ref{sum}). Although $\Gamma$ is also strongly
influenced by $\vec p_2$, this approximation is still adequate for
this parameter since $\sigma_2$ is smaller than $\sigma_3$ by an order
of magnitude. For example, from Table~\ref{table:map-fOGs} we infer
$\sigma_3 |p_{j3}| = 0.163, \, 0.038, \, 0.249$ for $j = 1,2,3$,
corresponding to $\Omegam, \Gamma, \sigma_8$, respectively. These
approximations are within 10\% of the MVB $\Delta \Theta_j$ for all
three parameters.

The eigenvectors of the Fisher matrix from the combined 2PCF $\xi_{\rm
  tot}$ are dominated by $\xi_+$. The $\xi_-$- and $\langle M_{\rm
  ap}^2 \rangle$-eigenvectors show similarities, which can be seen by
comparing $\vec p_3$ and $\vec p_4$ for the $(\Omegam, \Gamma,
\sigma_8, \Omega_\Lambda)$ cases (Tables \ref{table:map-OGsL} and
\ref{table:xi-OGsL}). Accordingly, the largest contribution to $\vec
p_3 (\vec p_4)$ comes from $\sigma_8 (\Omega_\Lambda$),
respectively, for both $\langle M_{\rm ap}^2 \rangle$ and $\xi_-$. In
the case of $\xi_{\rm tot}$ and $\xi_+$, the opposite is true.

For other combinations of parameters, the largest variance still makes
up more than 80\% of the total uncertainties and also dominates the
MVBs. Therefore, an optimisation scheme should try to maximise the
largest variance or smallest eigenvalue of the Fisher matrix. This
will be presented in the next section.

\subsection{Optimisation of the patch radius}
\label{sec:opt-patch}

Following KS04, we try to optimise our survey strategies by varying the number of lines of
sight per patch, $N$, and the patch radius, $R$, while keeping the total area constant.
Instead of focusing on the MVB for individual parameters as in that previous study, we
consider here the eigenvalues $\lambda_i$ of the Fisher matrix. In particular, we
concentrate on $\lambda_n$, the eigenvalue of the worst constrained principal component
since this dominates the MVB (see previous section). This is sufficient if only three
parameters are to be estimated from the data, since the variance of the first eigenvector
by far dominates the others. With four or more parameters, however, the
second principal component  has to be included in the
optimisation procedure.

The importance of the principal component corresponding to the smallest eigenvalue of
$\Mat F$ on the MVB can clearly be seen in Figs.~9--11 of KS04, where the MVB is plotted for
different $N$ and $R$. The curves are very similar for degenerate parameters, such as
$(\Omegam, \sigma_8)$ or $(\Gamma, n_{\rm s})$, since these pairs depend on the
eigenvectors of $\Mat F$ in a similar way. Moreover, in the case of a highly dominant
eigenvector as for the combination ($\Omegam, \sigma_8, \Gamma$), all three MVBs show the
same behaviour, since they all are dominated by this one principal component.

\begin{figure}
\protect\centerline{
\epsfysize = 3.3truein
\epsfbox[21 146 590 719]
{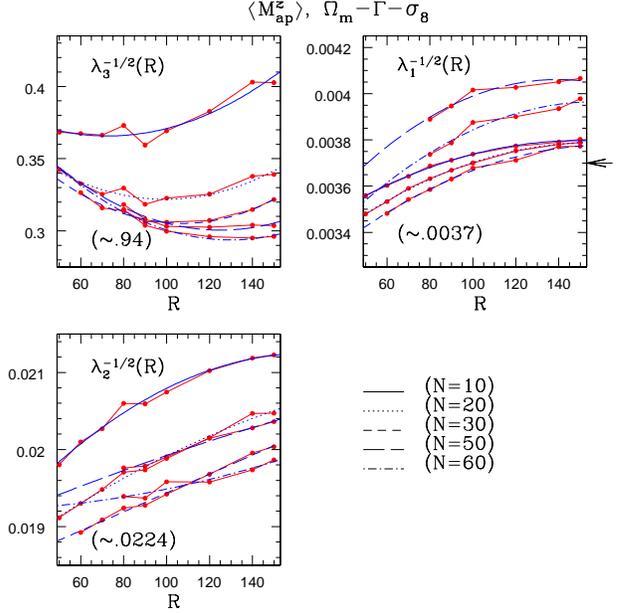}}
\caption{The variance $\sigma_i = \lambda_i^{-1/2}$ of the three eigenvectors
  of the Fisher matrix $\Mat F$, corresponding to ($\Omegam$,
  $\Gamma$, $\sigma_8$) and a flat Universe, for $\langle M_{\rm ap}^2
  \rangle$. Various surveys $(N, R)$ (Sect.~\ref{sec:survey}) are
  compared.  The results for the $300\cdot13^{\prime \, 2}$ survey is
  written in brackets in each panel and, while within the range of the
  plot, marked with an arrow. The smooth curves are second-order
  polynomial fits to the data points.  }
\label{fig:map1}
\end{figure}

\begin{figure}
\protect\centerline{
\epsfysize = 3.3truein
\epsfbox[21 146 590 719]
{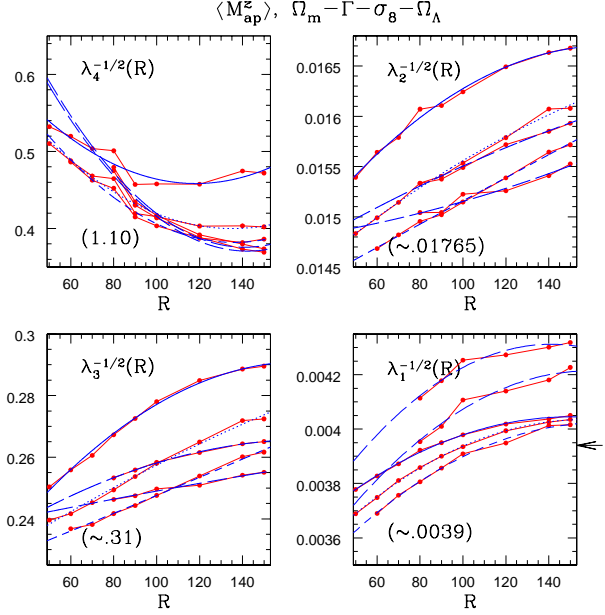}}
\caption{The variance $\sigma_i = \lambda_i^{-1/2}$ of the four eigenvectors
  of the Fisher matrix $\Mat F$, corresponding to ($\Omegam$,
  $\Gamma$, $\sigma_8, \Omega_\Lambda$), for $\langle M_{\rm ap}^2
  \rangle$. See Fig.~\ref{fig:map1} for more details.}
\label{fig:map2}
\end{figure}

\begin{figure}
\protect\centerline{
\epsfysize = 3.3truein
\epsfbox[21 146 590 719]
{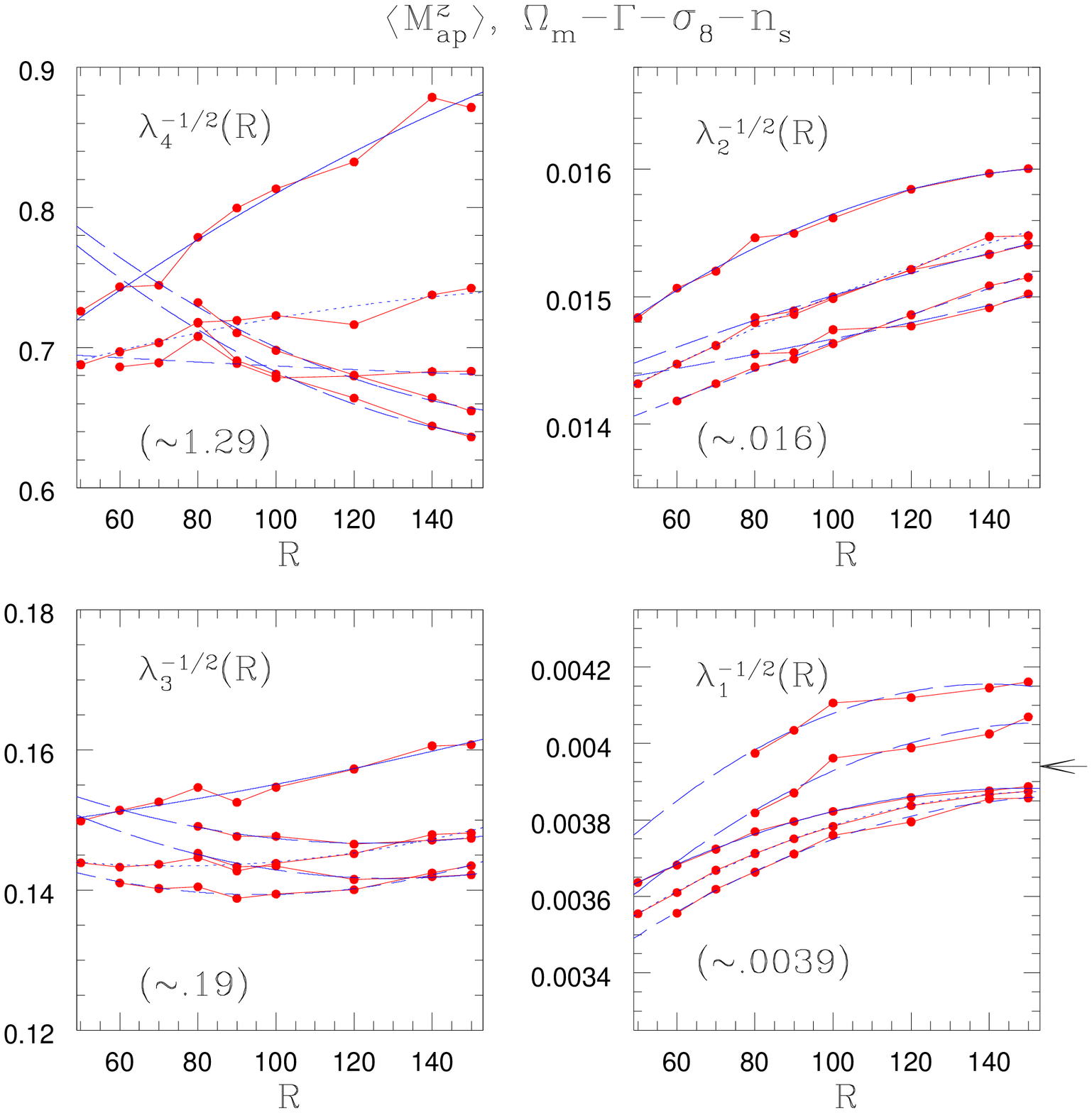}}
\caption{The variance $\sigma_i = \lambda_i^{-1/2}$ of the four eigenvectors
  of the Fisher matrix $\Mat F$, corresponding to ($\Omegam$,
  $\Gamma$, $\sigma_8, n_{\rm s}$) and a flat Universe, for $\langle M_{\rm ap}^2
  \rangle$. See Fig.~\ref{fig:map1} for more details.}
\label{fig:map3}
\end{figure}

\begin{figure}
\protect\centerline{
\epsfysize = 3.3truein
\epsfbox[21 146 590 719]
{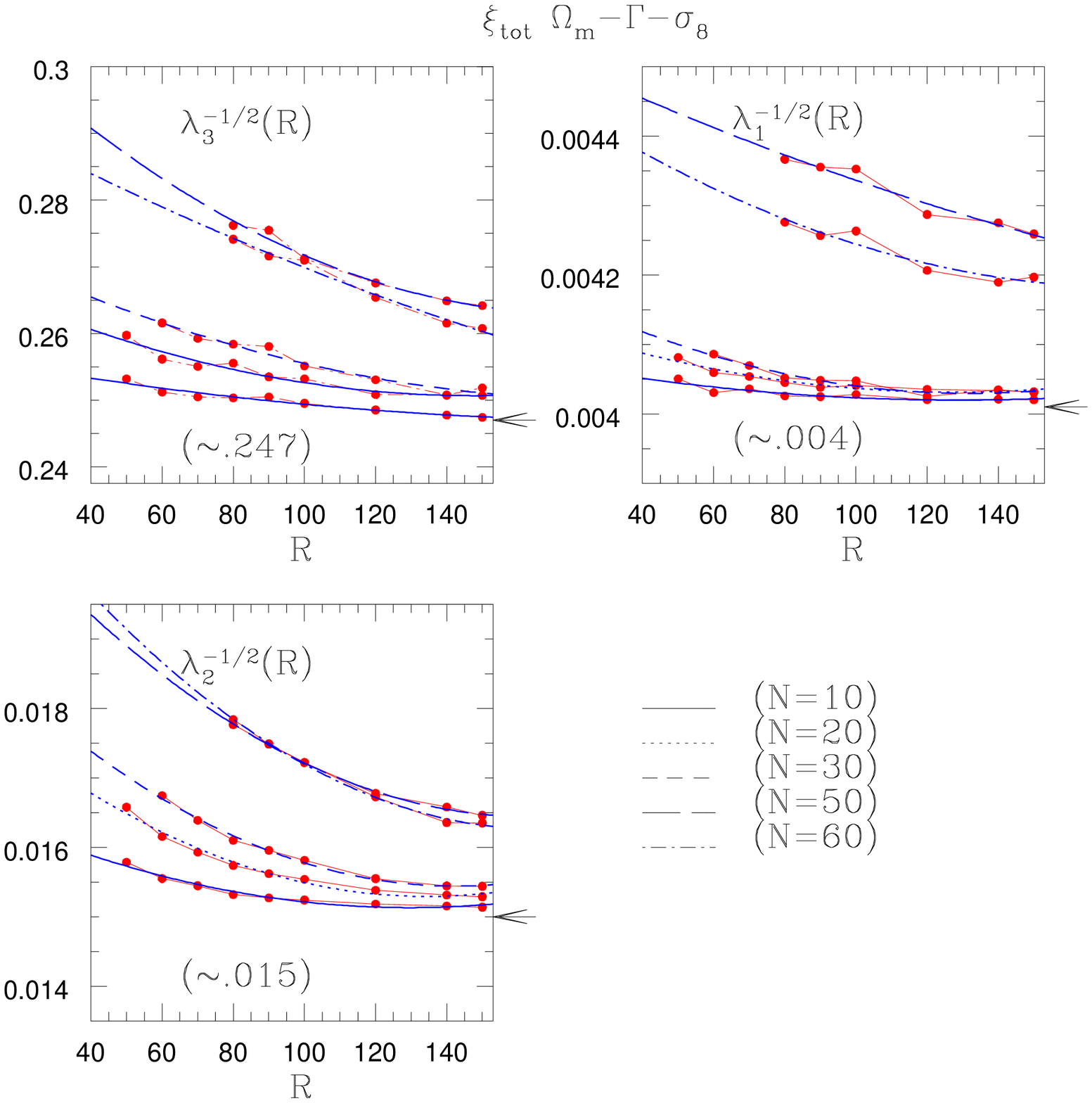}}
\caption{The variance $\sigma_i = \lambda_i^{-1/2}$ of the three eigenvectors
  of the Fisher matrix $\Mat F$, corresponding to ($\Omegam$,
  $\Gamma$, $\sigma_8$) and a flat Universe, for $\xi_{\rm
    tot}$. See Fig.~\ref{fig:map1} for more details.}
\label{fig:xi_1}
\end{figure}

\begin{figure}
\protect\centerline{
\epsfysize = 3.3truein
\epsfbox[21 146 590 719]
{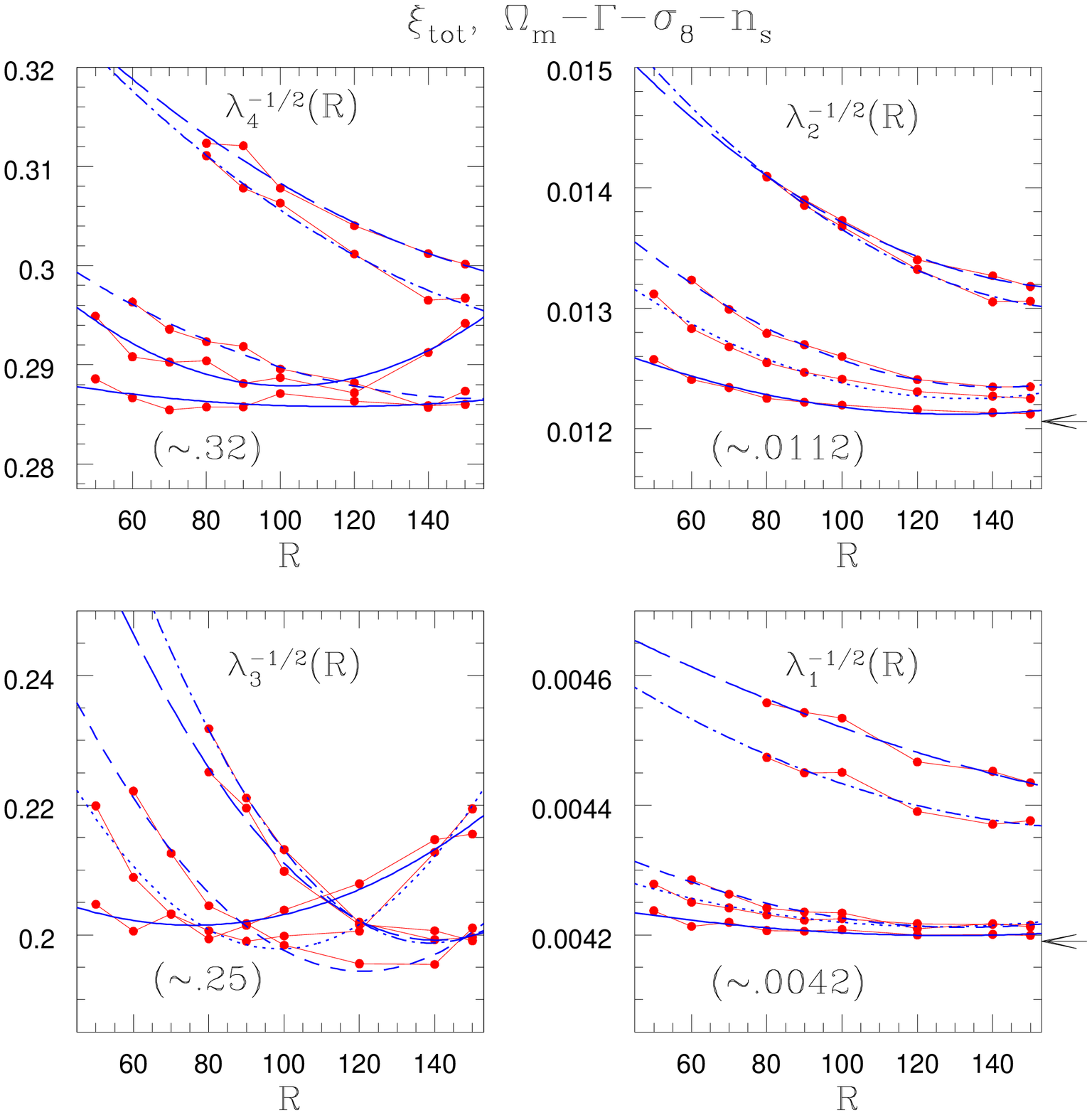}}
\caption{The variance $\sigma_i = \lambda_i^{-1/2}$ of the four eigenvectors
  of the Fisher matrix $\Mat F$, corresponding to ($\Omegam$,
  $\Gamma$, $\sigma_8, n_{\rm s}$) and a flat Universe, for $\xi_{\rm
    tot}$. See Fig.~\ref{fig:map1} for more details.}
\label{fig:xi_2}
\end{figure}

\begin{figure}
\protect\centerline{
\epsfysize = 3.3truein
\epsfbox[21 146 590 719]
{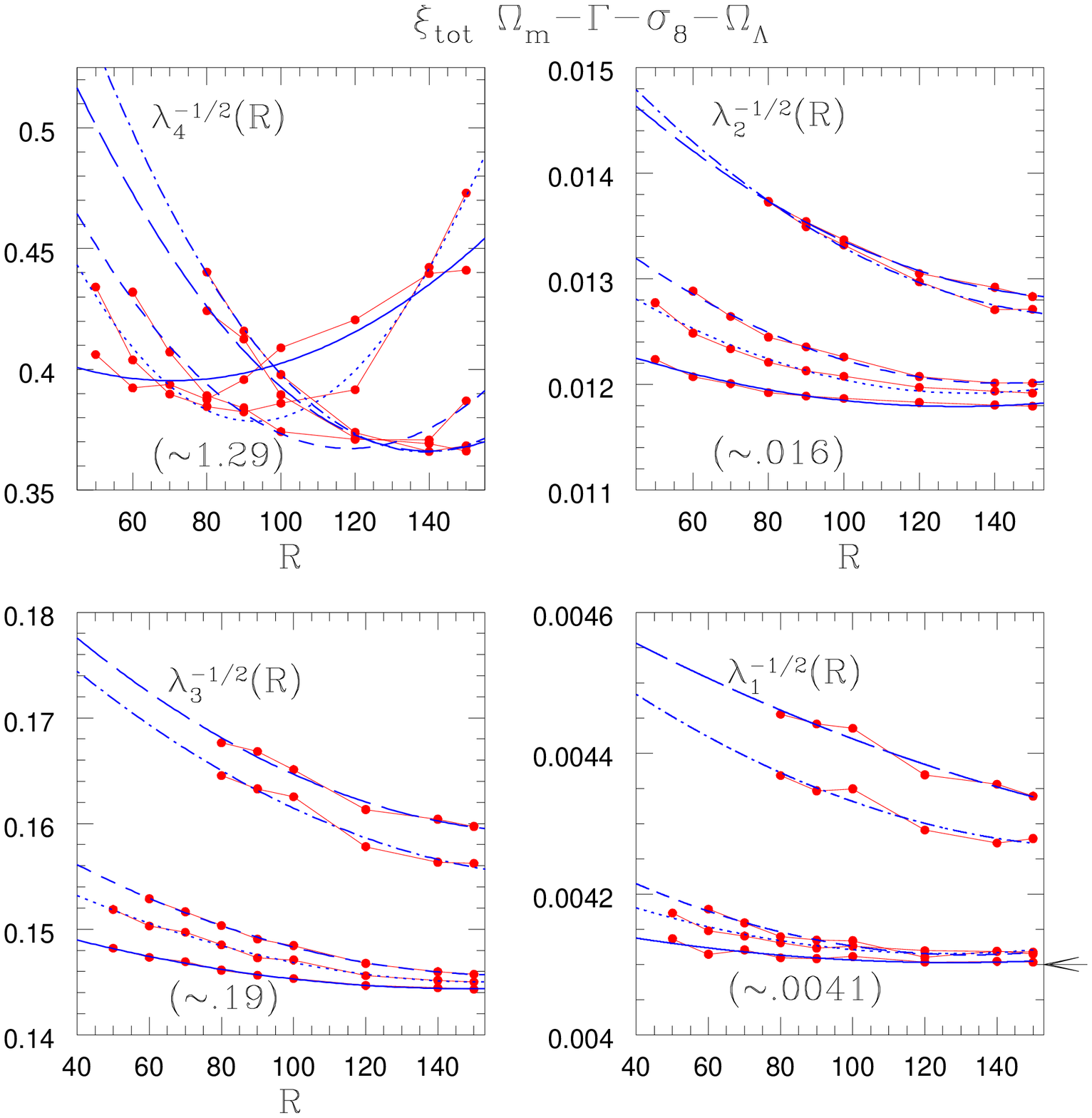}}
\caption{The variance $\sigma_i = \lambda_i^{-1/2}$ of the four eigenvectors
  of the Fisher matrix $\Mat F$, corresponding to ($\Omegam$,
  $\Gamma$, $\sigma_8, \Omega_\Lambda$), for $\xi_{\rm
    tot}$. See Fig.~\ref{fig:map1} for more details.}
\label{fig:xi_3}
\end{figure}

In Figs.~\ref{fig:map1}--\ref{fig:xi_3} we show the variance $\sigma_i =
1/\lambda_i^{-1/2}$ corresponding to the $i^{\rm th}$ eigenvector of the Fisher matrix.
The cosmological parameter combinations are the same than in the
previous section. We comment on the dependence on the patch radius $R$
and number of lines of sight per patch $N$ in the following sections.


\subsubsection{Aperture mass dispersion}
\label{sec:opt-map}

In the majority of the cases, the variance $\sigma_n$ of the
dominating, worst constrained parameter combination shows a minimum
for some $R_0$ within the probed range of patch radii. This confirms
the result of KS04, where the MVB (which is dominated by $\sigma_n$)
also showed a minimum. The optimal radius $R_0$ decreases towards
smaller $N$, thus, compact configurations yield better results than
sparse patches. This is also reflected in the fact that patches with
high $N$ are preferred over those with small $N$. For the other, less
dominant eigenvectors, the corresponding variance is (in most cases) a
monotonic increasing function of $R$.  All this implies that in order
to obtain constraints on parameter combinations, compact, densely
sampled patches will give the best results.

\subsubsection{Two-point correlation function}
\label{sec:opt-xi}

The radius $R_0$ where the variance $\sigma_n$ of the dominant
eigenvector takes a minimum is reached for larger $R$ than for
$\langle M_{\rm ap}^2 \rangle$. In some cases, $R_0$ exceeds 150 arc
minutes, which is the maximum of the probed range of radii.  In most
of the time, the variances $\sigma_i$ are decreasing functions of $R$
in contrast to the aperture mass dispersion. These results show that a
small cosmic variance is more important for $\xi_{\rm tot}$ than a
rigorous sampling of intermediate scales.  Strategies with sparse
patches probing a large number of independent regions on the sky and,
at the same time, capturing shear information on large angular scales
will provide the best constraints on cosmological parameters.

We compare the individual contributions of $\xi_+$ and $\xi_-$ to the
variances $\sigma_i$ of the joint estimator $\xi_{\rm tot}$. The
latter is dominated by $\xi_+$ and the dependence on the survey
strategy is very similar for both. The contribution from $\xi_{-}$,
however, resembles the one corresponding to the aperture mass
dispersion, since both statistics sample the convergence power
spectrum in a similar way. In particular, $\xi_-$ and $\langle M_{\rm
  ap}^2 \rangle$ do not probe large scales in contrast to $\xi_+$ (see
eq.~\ref{xi-Map-def}). The corresponding filter functions for
  the former two statistics, $\J_4(\eta)$ and $[\J_4(\eta)/\eta^2]^2$,
  respectively, are both proportional to $\eta^4$ for small $\eta$,
  suppressing large scales. On the contrary, for the latter
  statistics, $\J_0$ is constant for small arguments. A survey which
  covers large angular scales at the expense of a dense sampling of
  small and medium scales will therefore not be optimal for $\xi_-$
  and $\langle M_{\rm ap}^2 \rangle$, but efficient for $\xi_+$.

\subsubsection{Comparison with the uncorrelated lines-of-sight-survey}
\label{sec:opt-comp}

The $300\cdot13^{\prime \, 2}$-survey of uncorrelated lines of sight
shows larger values for $\sigma_i,\, i=2\ldots n$ than the patch
surveys, corresponding to poorer constraints on cosmological
parameters. However, the smallest variance $\sigma_1$ is reached
asymptotically for patch strategies with large $R$ and $N=10,20,30$.
These surveys consisting of a large number of sparse patches
corresponding to a small cosmic variance are those which are most
similar to the uncorrelated lines-of-sight-survey. Since for the 2PCF
a small cosmic variance is crucial, the patch strategies show not much
improvement in $\sigma_1$ over the $300\cdot13^{\prime \, 2}$-survey (unless both
$\Omegam$ and $\Omega_\Lambda$ are free parameters, see
Fig.~\ref{fig:xi_3}). This is in contrast to the aperture mass
statistics, where the improvement in $\sigma_1$ is more than a factor
of two.

With decreasing $i$, the variance $\sigma_i$ of the eigenvector
$\vec p_i$ becomes less sensitive on the survey geometry. The
measurement of the best constrained combinations of parameters can
therefore not efficiently be improved and even the $300\cdot13^{\prime
  \, 2}$-survey will yield good results.

The $300\cdot13^{\prime \, 2}$-survey can compete with a patch
strategy regarding the best constrained eigenvectors which contribute
least to the parameter uncertainties. For the dominant parameter
combinations, the patch strategies are superior and yield much better
constraints on cosmological parameters.

\subsection{Inclusion of additional parameters}
\label{sec:bigger}

To include more cosmological parameters in our analysis, we consider
a survey of 141 square degree area. This corresponds to an enlargement
of the $(30,100^\prime)$ strategy, consisting of 100 instead of 10
independent patches. The eigenvectors and corresponding eigenvalues
are shown in Table \ref{tab:xi-141}.

The first, best determined eigenvector is orthogonal to the
$\Omegam$-$\sigma_8$ degeneracy direction, as in
Sect.~\ref{sec:eigenvec}. The second best direction is orthogonal to
the prominent $\Gamma$-$n_{\rm s}$ degeneracy direction, but with a
strong contribution from $\Omegam$. The worst constrained eigenvector
$\vec p_7$ is almost solely dependent on the redshift parameters $z_0$ and $\beta$.

\begin{table*}
  \caption{Eigenvectors of the Fisher matrix for all 7 parameters from
    a 141 square degree survey.}
  \label{tab:xi-141}
  \begin{center}
    \begin{tabular} {@{}lc|ccccccc|c}
\hline
\hline
& &  $\vec p_1$ & $\vec p_2$ & $\vec p_3$ & $\vec p_4$ & $\vec p_5$ & $\vec p_6$ & $\vec p_7$ &  $\triangle \Theta_i$ \\ \hline
& $\Omegam$     & $\mi 0.639$ &   $-0.563$    &   $-0.016$    &   $\mi 0.260$ &   $\mi 0.002$ &   $-0.454$ &      $\mi 0.014$  &  $0.622$ \\
& $\Gamma$      & $\mi 0.392$ &   $\mi 0.691$ &   $-0.404$    &   $\mi 0.219$ &   $-0.360$    &   $-0.168$ &      $\mi 0.002$      &      $0.233$ \\
& $\sigma_8$    & $\mi 0.503$ &   $-0.049$    &   $-0.359$    &   $-0.556$    &   $\mi 0.301$ &   $\mi 0.463$  &  $-0.034$ &      $0.669$ \\
& $z_0$         & $\mi 0.311$ &   $\mi 0.088$ &   $\mi 0.580$ &   $-0.074$    &   $-0.317$    &   $\mi 0.283$  &  $\mi 0.611$      &      $4.099$ \\
& $n_{\rm s}$   & $\mi 0.176$ &   $\mi 0.335$ &   $\mi 0.266$ &   $\mi 0.437$ &   $\mi 0.767$ &   $\mi 0.078$  &  $\mi 0.025$      &      $0.218$ \\
& $\beta$       & $-0.236$    &   $-0.071$    &   $-0.469$    &   $\mi 0.011$ &   $\mi 0.236$ &   $-0.196$ &      $\mi 0.791$      &      $5.292$ \\
& $\Omega_\Lambda$  & $-0.038$&   $-0.279$    &   $-0.285$    &   $\mi 0.615$ &   $-0.188$    &   $\mi 0.653$  &  $\mi 0.004$      &      $0.885$ \\ \hline
& $\sigma_i$    & $\mi 0.001$ &   $\mi 0.004$ &   $\mi 0.033$ &   $\mi 0.063$ &   $\mi 0.117$ &   $\mi 1.354$  &  $\mi 6.685$      &              \\
\hline\hline
\end{tabular}
  \end{center}
\end{table*}

\subsection{Local degeneracy directions}
\label{power-law}

\begin{figure}
  \resizebox{1.0\hsize}{!}{ 
     \includegraphics{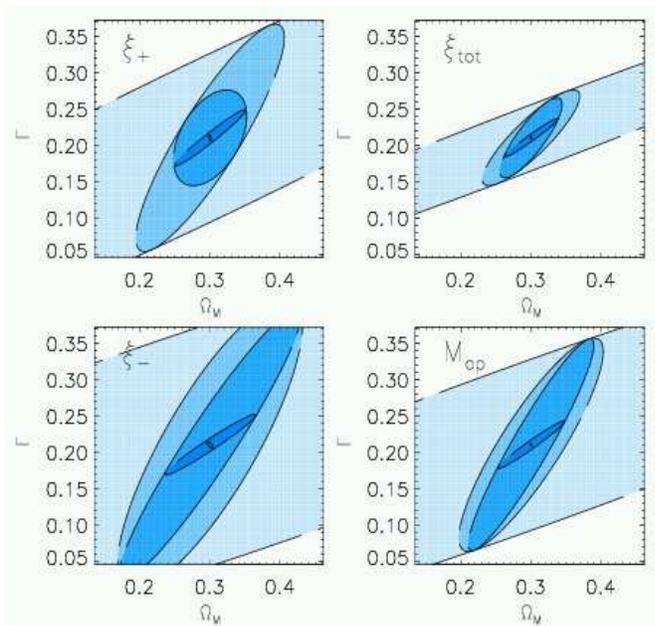}
  }
  \caption{1$\sigma$ error ellipses in the $\Omegam$-$\Gamma$ plane.
    From small to large ellipses, successive marginalisation over
    $\sigma_8, n_{\rm s}, \Omega_\Lambda, \beta, z_0$ (except the ones
    that are plotted) was performed.  The four panels correspond to
    the four estimators $\xi_+$, $\xi_-$, $\xi_{\rm tot}$ and $\langle
    M_{\rm ap}^2 \rangle$ as indicated.}
  \label{fig:comp_deg}
\end{figure}

\begin{figure}
  \resizebox{1.0\hsize}{!}{
    \includegraphics{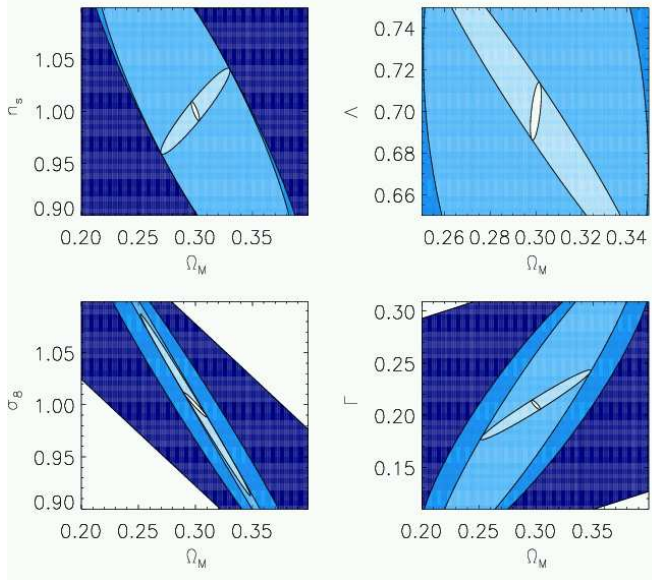}
  }
  \caption{1$\sigma$ error ellipses for $\Omegam$ and $\sigma_8$,
    $n_{\rm s}$, $\Omega_\Lambda$ and $\Gamma$, respectively, in the
    four panels. From small to large ellipses, successive
    marginalisation over the remaining parameters out of $(\Gamma,
    \sigma_8, n_{\rm s}, \Omega_\Lambda, \beta, z_0)$ was performed.
    The estimator is $\langle M_{\rm ap}^2 \rangle$ and the survey
    strategy the 141 square degree survey.}
  \label{fig:params}
\end{figure}

From the Fisher matrix, we quantify the local direction of degeneracy
between the parameter pairs $(\Omegam, \sigma_8)$ and $(\Gamma, n_{\rm
  s})$. For each pair, we marginalise over the remaining parameters
out of  $(\Omegam, \sigma_8, \Gamma$, $n_{\rm
  s})$ for a flat Universe, and find
\begin{equation}
  \sigma_8 \, \Omegam^{0.48} = 0.56 \quad \mbox{and} \quad n_{\rm s}
  \, \Gamma^{0.3} = 0.6,
\end{equation}
assuming a power-law dependence between parameter pairs which is usually found
in likelihood analysis. These numerical coefficients are basically the same
for $\xi_{\rm tot}$ and $\langle M_{\rm ap}^2 \rangle$. The resulting
degeneracy directions are in agreement with Kilbinger \& Schneider (2005),
even though a non-Gaussian shear field was used to calculate the covariance in
this previous work.  Marginalisation over the hidden parameters increases the
volume of the error ellipsoid and also alters the orientation of its axes in
parameter space. However, the general degeneracy direction is similar for
various marginalised parameters, see Figs.~\ref{fig:comp_deg} and
\ref{fig:params}. We conclude that the directions of near-degeneracy between
the considered parameter pairs are robust against the inclusion of
non-Gaussianity of the shear field, but less stable against the addition of
cosmological parameters. In Fig.~\ref{fig:comp_deg} we show the
1$\sigma$-ellipses for the parameter pair ($\Omegam$, $\Gamma$) using all four
estimators $\xi_+$, $\xi_-$, $\xi_{\rm tot}$ and $\langle M_{\rm ap}^2
\rangle$. Slight misalignments in the degeneracy directions from $\xi_+$
compared with $\xi_-$ leads to improved constraints from the combined
estimator $\xi_{\rm tot}$.


\subsection{Effect of the survey depth on the principal components}
\label{sec:depth}

\begin{figure}[!tb]
  \resizebox{1.0\hsize}{!}{
    \includegraphics[bb=35 144 565 460]{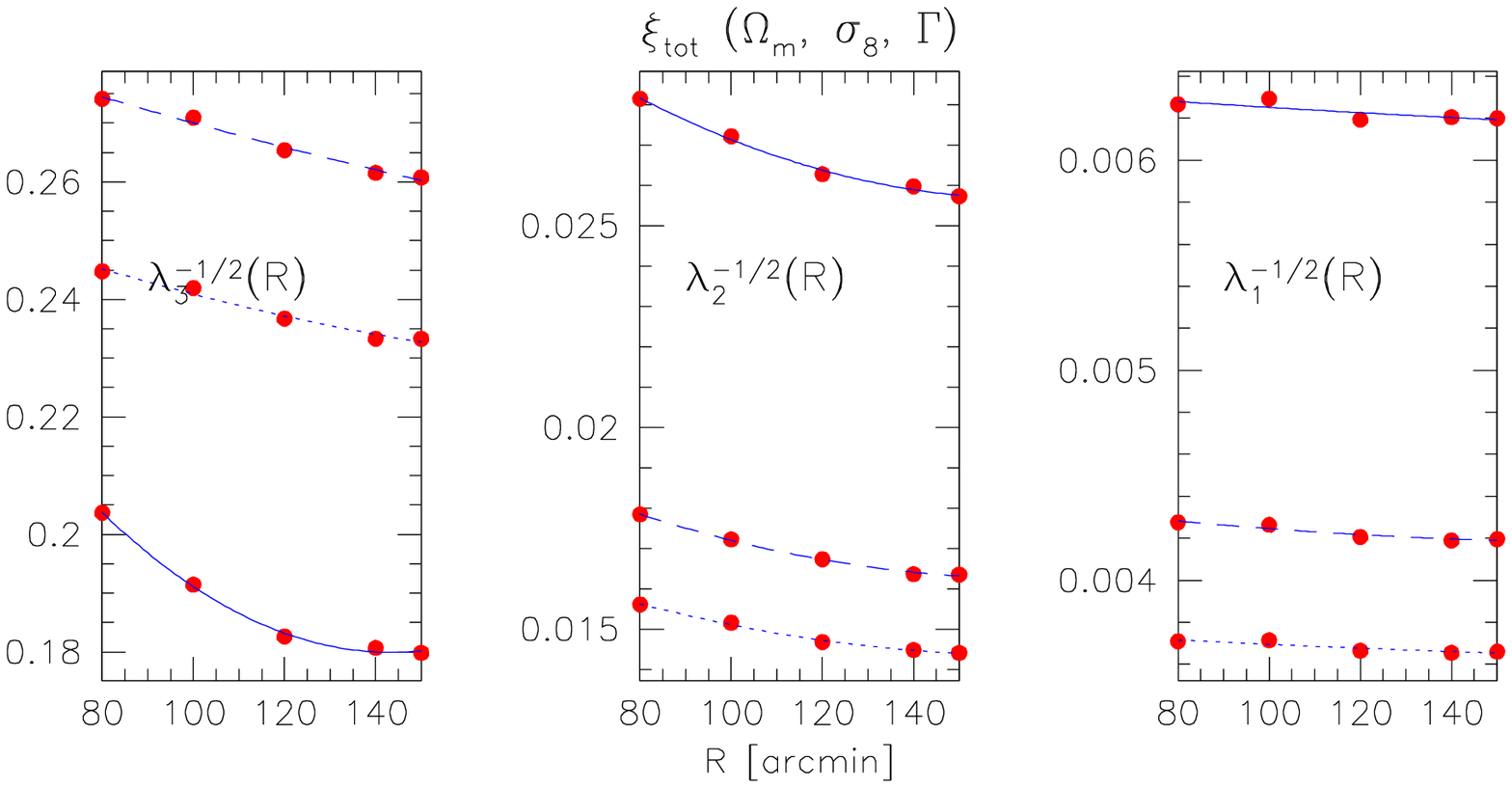}
  }
  \resizebox{1.0\hsize}{!}{
    \includegraphics[bb=35 144 565 450]{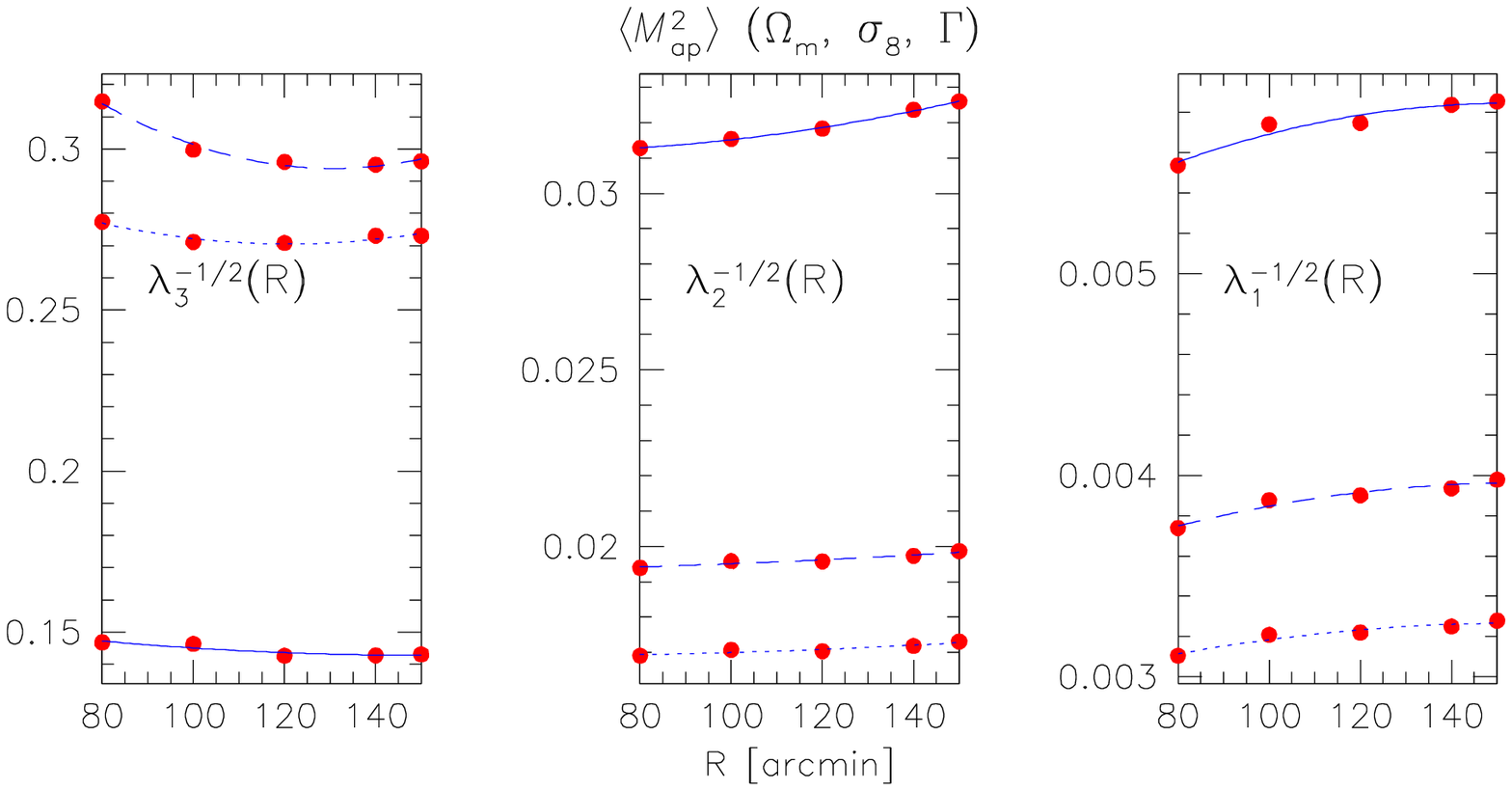}
  }
  \caption{The variance $\sigma_i = \lambda_i^{-1/2}$ of the three
    eigenvectors of the Fisher matrix corresponding to $(\Omegam,
    \sigma_8, \Gamma)$ and a flat Universe, using the 2PCF (top) and
    the aperture mass dispersion (lower panels). The curves show
    different survey depth where solid, dashed and dotted lines
    correspond to $z_0 = 0.8, 1.0$ and 1.2, respectively.}
  \label{fig:lambda_OGs_4_z}
\end{figure}

For surveys with fixed $N=60$ and varying $R$, we calculate the Fisher
matrix for three different source redshift distribution functions,
parametrised by $z_0 = 0.8, 1.0$ and 1.2, respectively. The variance
$\sigma_i$ of the principal components of the Fisher matrix
corresponding to the three parameters ($\Omegam, \sigma_8$, $\Gamma$)
is shown in Fig.~\ref{fig:lambda_OGs_4_z}. The general characteristics
of $\lambda_i^{-1/2}$ as a function of $R$ does not change with the
survey depth and simply gets scaled. The variance corresponding to
the first two eigenvectors decrease with increasing survey depth,
corresponding to a smaller error on the parameter measurement. However
and unexpectedly, the principal component corresponding to the largest
variance $\sigma_3$ is best determined for the most shallow survey
with $z_0 = 0.8$. This principal component takes little contribution
from $\Gamma$ and points in the $\Omegam$-$\sigma_8$ degeneracy
direction. We repeat the PCA without the shape parameter and found a
similar result. Also when marginalising over additional parameters,
the largest eigenvalue of the Fisher matrix is smallest for $z_0 =
0.8$.

This unexpected effect is a local one, present in the Fisher matrix
only. In order to see the global behaviour, we calculate the
likelihood in the $\Omegam$-$\sigma_8$ plane and find that the confidence
levels get tighter with increasing source redshift.
The near-degeneracy is less pronounced if the survey is deeper and
more information about the large-scale structure is collected in the
shear signal. Moreover, the curvature of confidence levels increases
with increasing $z_0$ which leads to a larger bending of the curves of
constant likelihood. The shear correlation from different source
galaxy redshift distributions allows one to constrain slightly
different regions in parameter space. This fact is made use of in
shear tomography to lift the parameter near-degeneracies (Hu 1999).

\subsection{Effect of the source galaxy density on the principal components}
\label{sec:ngal}

\begin{figure}[!tb]
  \resizebox{1.\hsize}{!}{
    \includegraphics[bb=35 144 565 450]{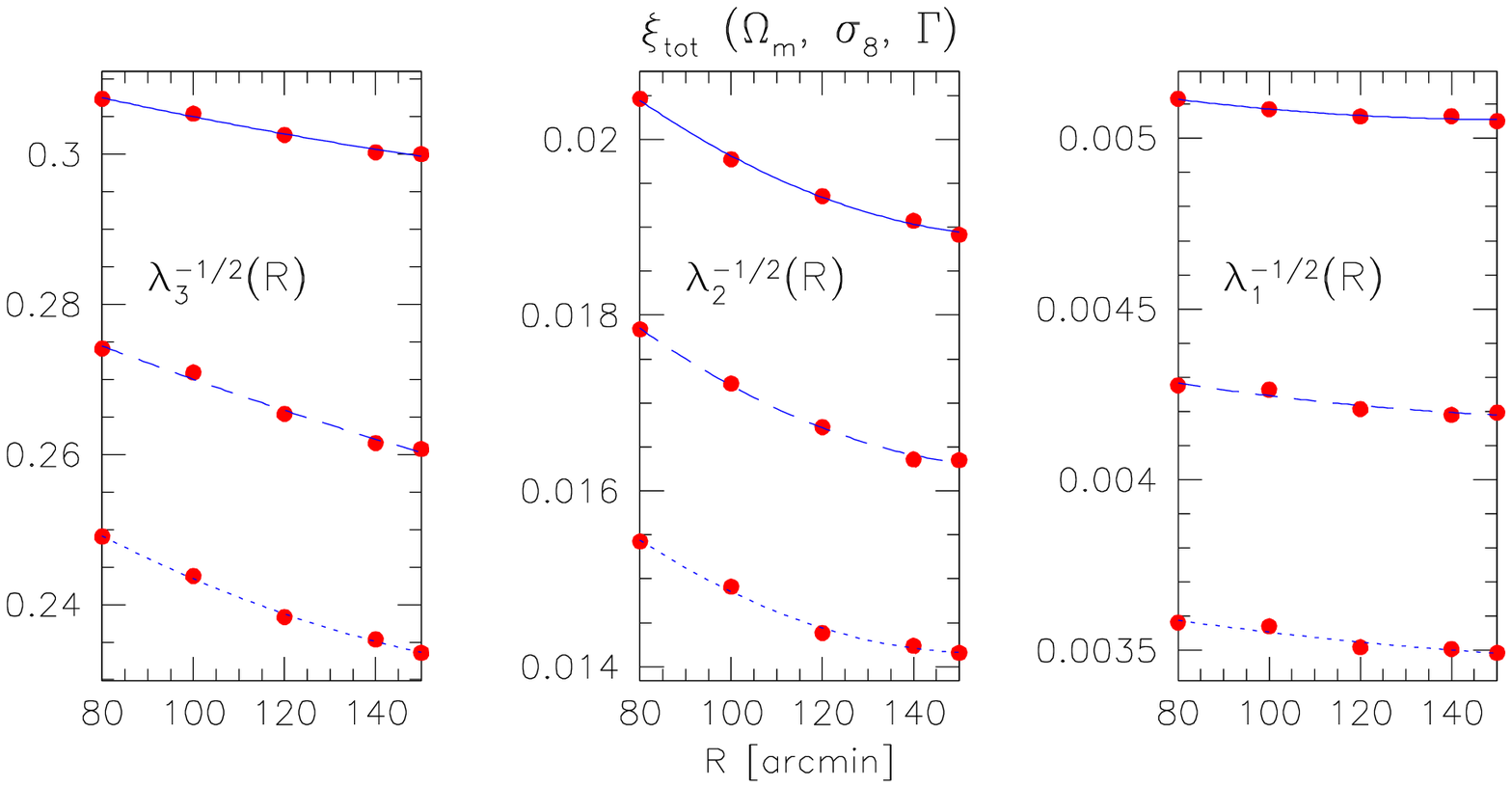}
  }
  \resizebox{1.0\hsize}{!}{
    \includegraphics[bb=35 144 565 450]{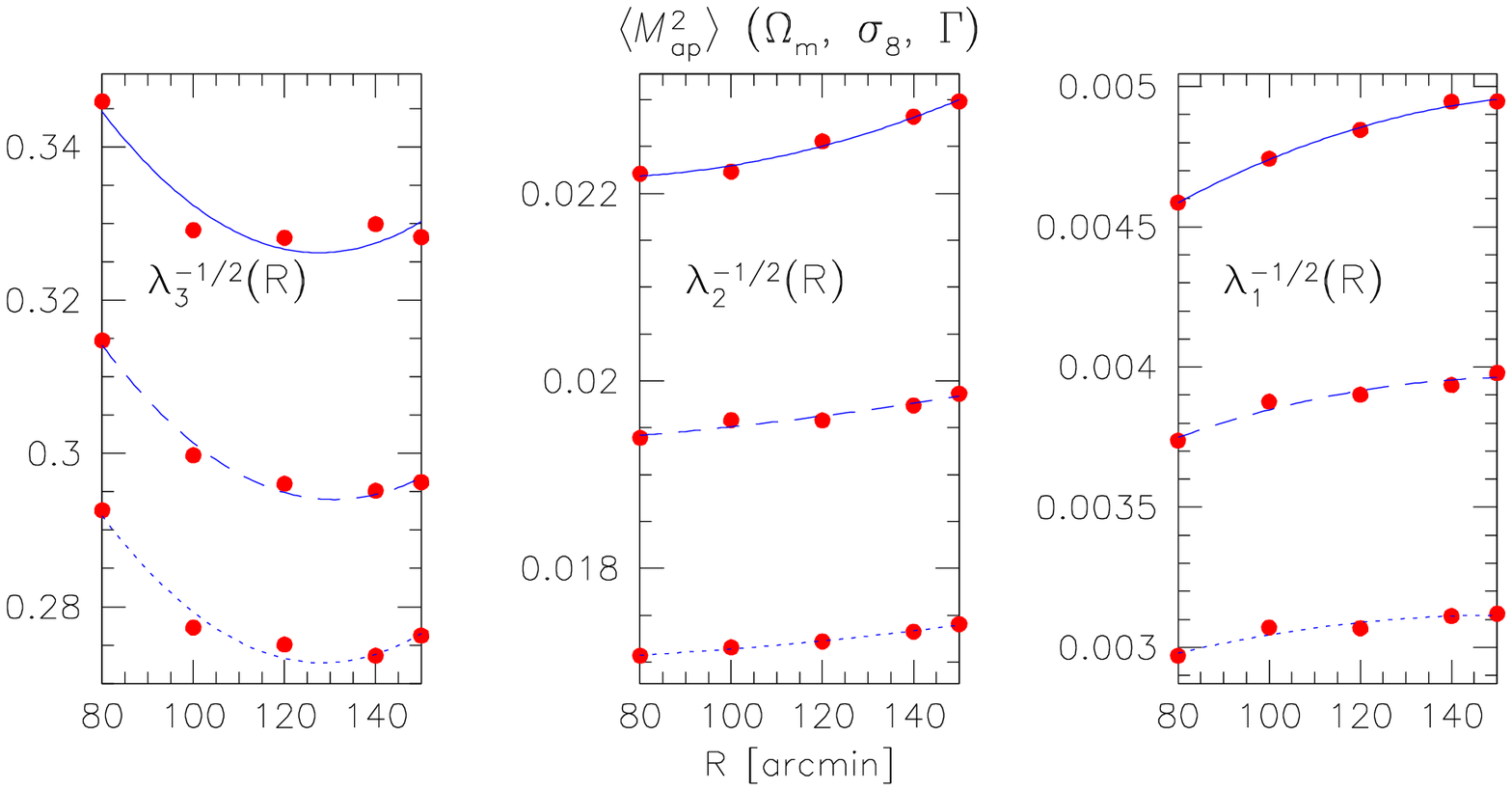}
  }
  \caption{The variance $\sigma_i = \lambda_i^{-1/2}$ of the three
    eigenvectors for the Fisher matrix corresponding to $(\Omegam,
    \sigma_8, \Gamma)$ and a flat Universe, using the 2PCF (top) and the
    aperture mass dispersion (lower panels). The curves show different
    survey depth where solid, dashed and dotted lines correspond to
    $n_{\rm gal} = 20, 30$ and 40 per square arc minute,
    respectively.}
  \label{fig:lambda_OGs_4_n}
\end{figure}

For the same survey types ($N=60$) and the parameters $\Omegam,
\sigma_8$ and $\Gamma$ as in the previous section, we compute the
Fisher matrix for three different surface densities of background
galaxies, $n_{\rm gal} = 20, 30$ and 40 per square arc minute. The
variance $\sigma_i$ of the three principal components of the Fisher
matrix is plotted as a function of patch radius $R$ in
Fig.~\ref{fig:lambda_OGs_4_n}, corresponding to measurements of the
2PCF and the aperture mass dispersion, respectively. As expected,
$\sigma_i$ decreases with increasing $n_{\rm gal}$ since more
background galaxies provide a better sampling of the shear field. The
minimum variance for $\sigma_3$ at $R_0 \approx 130$ in the case of
$\langle M_{\rm ap}^2 \rangle$ does not change with noise level.

In this work, we treated the survey depth, parametrised by $z_0$,
and the number density of background galaxies $n_{\rm gal}$ as
independent survey characteristics. For a realistic survey however,
this assumption is not justified in general since an increase in
depth implies a higher galaxy number density.

\subsection{Principal components and the error ellipsoid volume}
\label{sec:volume}

We investigate the geometric mean $\rho =
(\lambda_1\dots\lambda_n)^{1/n}$ of the $n$ eigenvalues of the Fisher
matrix $\Mat F$. The equivalent radius of the Fisher error ellipsoid,
which is the radius of an $n$-sphere with the same volume, is
$\rho^{-1/2} = (\det {\Mat F})^{-1/2n}$. 

\begin{figure}
  \protect\centerline{
    \epsfysize = 3.5truein
    \epsfbox[22 147 590 719]
    {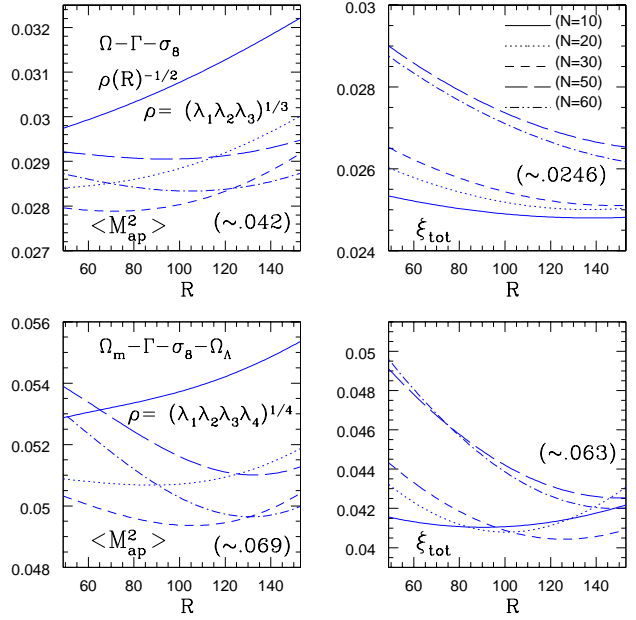}}
  \caption{The effective radius $\rho^{-1/2}$ of the Fisher error ellipsoid
    as a function of survey radius $R$. The left panels
    correspond to $\langle M_{\rm ap}^2 \rangle$, the right panel to $\xi_{\rm tot}$.
    In the upper two panels, the parameter combination is $(\Omegam,
    \Gamma, \sigma_8)$, in the lower panels it is $(\Omegam, \Gamma,
    \sigma_8, \Omega_{\Lambda})$. The numbers in parenthesis
    correspond to $\rho^{-1/2}$ from the $300\cdot13^{\prime \,
      2}$-survey. The definition of $\rho$ is given in the panel as a reminder.}
  \label{fig:volume1}
\end{figure}

\begin{figure}
  \protect\centerline{
    \epsfysize = 1.9truein
    \epsfbox[28 425 590 719]
    {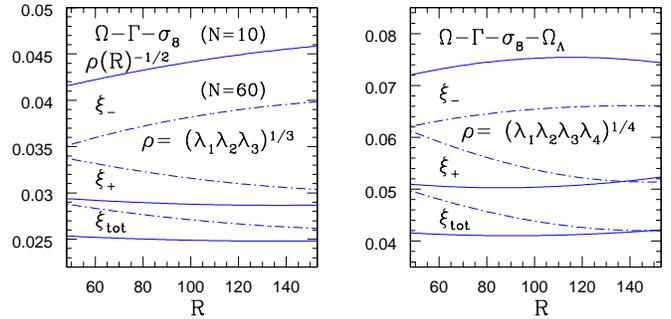}}
  \caption{The effective radius $\rho^{-1/2}$ of the Fisher error
    ellipsoid corresponding to ($\Omegam, \Gamma, \sigma_8)$ (left
    panel) and $(\Omegam, \Gamma, \sigma_8, \Omega_{\Lambda})$ (right
    panel). The three estimators $\xi_-$, $\xi_+$ and $\xi_{\rm tot}$
    are compared using for surveys with sparse ($N=10$, solid lines)
    and dense ($N=60$, dash-dotted lines) geometries.}
\label{fig:volume2}
\end{figure}

In Figs.~\ref{fig:volume1} and \ref{fig:volume2} we plot the effective
error ellipse radius $\rho^{-1/2}$ as a function of $R$ for survey
geometries $(N,R)$ using different parameter combinations and
estimators. Since $\rho^{-1/2}$ is dominated by the largest variance
$\sigma_n$ (smallest eigenvalue of $\Mat F$), it shows a similar
behaviour (see Sect.~\ref{sec:opt-patch}). However, the minimum seems
to be reached at smaller $R_0$ an therefore the optimal survey radius
is smaller  than the previous sections implied. The
$300\cdot13^{\prime \, 2}$-survey (see Fig.~\ref{fig:volume1} and
Table \ref{tab:rho300}) yields comparable results than for the patch
geometries if $\xi_+$ is used. $\xi_-$ and $\langle M_{\rm ap}^2
\rangle$ (not shown), however, strongly suffer from the lack of scales
larger than 20 arc minutes which results in a much larger error
ellipsoid.

\begin{table}[b]
  \caption{The effective radius of the Fisher error ellipse
    $\rho^{-1/2}$ for the $300\cdot13^{\prime \, 2}$-survey,
    corresponding to two combinations of cosmological parameters. The
    values of $\rho^{-1/2}$ for various patch strategies $(N,R)$ are
    displayed in Fig.~\ref{fig:volume2}.}
  \label{tab:rho300}
  \begin{center}
    \begin{tabular}{l|cc} \hline\hline
      & $(\Omegam, \Gamma, \sigma_8)$ & $(\Omegam, \Gamma, \sigma_8,
      \Omega_{\Lambda})$ \\ \hline
      $\xi_{\rm tot}$ & 0.024 & 0.061 \\
      $\xi_+$         & 0.028 & 0.062 \\
      $\xi_-$         & 0.130 & 0.163 \\ \hline\hline
    \end{tabular}
  \end{center}
\end{table}

In Fig.~\ref{fig:volume2}, the results for the 2PCFs $\xi_+$ and
$\xi_-$ are compared. As already mentioned before, $\xi_{\rm tot}$ is
dominated by $\xi_+$ and shows a similar behaviour. The shape of the
$\xi_-$-curves on the other hand resembles the one for $\langle M_{\rm
  ap}^2 \rangle$.

\subsection{Principal components of the scaled Fisher matrix}
\label{sec:scaled}

\begin{figure}
\protect\centerline{
\epsfysize = 1.9truein
\epsfbox[30 430 590 719]
{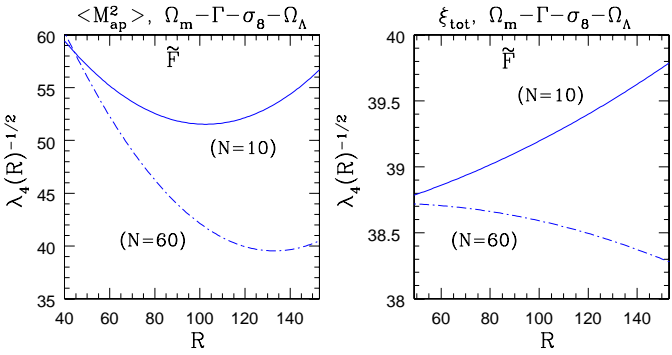}}
\caption{The variance $\sigma_4 = \lambda_4^{-1/2}$ of the dominant
  eigenvalue of the scaled Fisher matrix $\tilde F$ for the parameters
  ($\Omegam, \Gamma, \sigma_8, \Omega_\Lambda$), as a function of the
  patch radius $R$. Solid lines correspond to surveys with $N=10$,
  dash-dotted curves are for $N=60$.  }
\label{fig:scale}
\end{figure}

The scaled Fisher matrix $\tilde F$ is defined as
\begin{equation}
\tilde F_{ij} = {F_{ij} \over \sqrt{F_{ii} F_{jj}} }.
\end{equation}
Its inverse $\tilde {\Mat F}^{-1}$ represents the covariance of the
normalised (unit variance), correlated Gaussian variables $\tilde
\Theta_j$ constructed from the original cosmological parameters,
$\tilde \Theta_j = \Theta_j / \sqrt{\langle \Theta_j^2\rangle}$. The
scaled parameters can be compared with each other in a 
straightforward way, and their correlation gets more underlined.  The
variation with patch radius $R$ of the eigenvalues of the scaled
Fisher matrix matches well with their unscaled counterparts. In
Fig.~\ref{fig:scale} we plot the variance of the dominant eigenvector
$\lambda_n^{-1/2}(R)$ for geometries with $N=10$ and $N=60$,
respectively, and the cosmological parameters $(\Omegam, \Gamma,
\sigma_8, \Omega_{\Lambda})$.

\subsection{Principal components and the use of priors}
\label{sec:priors}

\begin{figure}
\protect\centerline{
\epsfysize = 1.9truein
\epsfbox[30 430 590 719]
{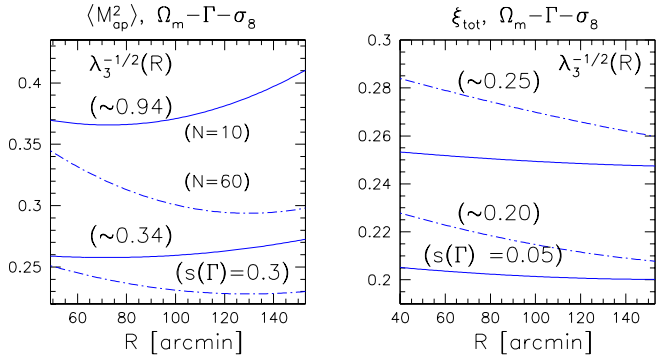}}
\caption{The variance of the dominant eigenvector $\sigma_3$ is plotted for patch
  geometries with $N=10$ (solid lines) and $N=60$ (dash-dotted), for
  $\langle M_{\rm ap}^2 \rangle$ (left panel) and $\xi_{\rm tot}$
  (right panel).  The three cosmological parameters under
  consideration are $\Omegam$, $\Gamma$ and $\sigma_8$. The upper
  curves correspond to the case without prior. For the lower pair of
  curves, and additional prior on $\Gamma$ of $s(\Gamma) = 0.3 \,
  (0.05)$ is assumed for $\langle M_{\rm ap}^2 \rangle \; (\xi_{\rm
    tot})$, respectively. The values in the parenthesis denote $\sigma_n$ the
  $300\cdot13^{\prime \, 2}$-survey.
}
\label{fig:prior}
\end{figure}

Additional priors on parameters modify the original Fisher matrix $F$
to $\hat F = F + C^{-1}$. For example, a Gaussian prior for the $i^{\rm
  th}$ parameter with variance $s_i$ corresponds to $C_{ij}^{-1} =
\delta_{ij} \, s_i^{-2}$. Priors lower the eigenvalues.  Since the
reduction is the larger the smaller the eigenvalue is, the dominant
eigenvector with the smallest eigenvalue is affected most by priors.
The effect of a prior on the variance of this eigenvector is shown in
the upper right panel of Fig.~\ref{fig:lambda_scale} and in
Fig.~\ref{fig:prior}. In Fig.~\ref{fig:lambda_scale}, the priors
$s(p_i)$ are 0.003 for $\Omegam, \sigma_8$ and $\Gamma$, 0.03
for $\Omega_\Lambda$, 0.1 for $n_{\rm s}$ and $z_0$ and 1 for
$\beta_p$. For the three curves, priors are added for parameters as
indicated in the panel. The more priors are included, the more is the
large, dominant variance affected, and the curve $\sigma_i$ as a
function of $i$ becomes less steep.

\begin{figure}[!tb]
  
  \resizebox{\hsize}{!}{
    \includegraphics[bb=38 444 572 698]{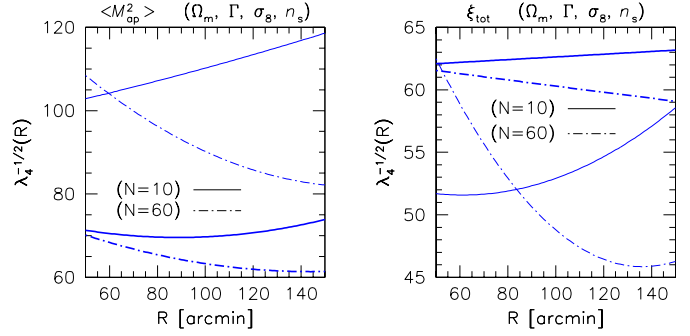}
  }
  \caption{The variance of the dominant eigenvector $\sigma_4 =
    \lambda_4^{-1/2}$ of the scaled Fisher matrix $\tilde F$ for the
    parameters $(\Omegam, \Gamma, \sigma_8, n_{\rm s}$), as a function
    of the patch radius $R$. Thick curves are for a flat Universe,
    thin curves for constant $\Omega_\Lambda$. Solid lines correspond
    to $N=10$, dash-dotted lines to $N=60$.}
  \label{fig:flat_prior}
\end{figure}

The addition of a prior also flattens the curve $\sigma_n(R)$ as a function of
patch radius $R$ and the difference between the surveys becomes less
pronounced, making the optimisation less effective, see Fig.~\ref{fig:prior}.
From Fig.~\ref{fig:flat_prior} one infers that the prior of a flat Universe
($\Omega_\Lambda = 1 - \Omegam$) corresponds to a higher variance $\sigma_n$
than a constant $\Omega_\Lambda = 0.7$ prior.

\section{Summary and discussion} 
\label{sec:discussion}

We study the design of cosmic shear survey with the means of a
principal component analysis of the Fisher information matrix.
Although only a local approximation of the likelihood, the Fisher
matrix provides a simple and cost-effective method to predict
cosmological parameter constraints and to investigate degeneracies
between parameters. We compare realistic survey designs with
non-trivial topology. In most of the previous analyses (an exception
being KS04) either simple monolithic survey topologies were assumed or
the correlation of power for different Fourier modes was neglected. In
this work, we use the full covariance matrix of second-order shear
statistics corresponding to a complicated, non-trivial distribution of
lines of sight. We assume the shear field to be Gaussian which leads
to an underestimation of the covariance on intermediate angular scales
between $\sim$ 1 and 10 arc minutes.

The eigenvectors or principal components of the Fisher matrix
determine linear, orthogonal combinations of cosmological parameters.
We find that these vectors are only little affected by the survey
characteristics. The same is true for the directions of
near-degeneracy between parameters. However, the eigenvalues change
significantly with survey geometry. The eigenvector of the Fisher
matrix which dominates the errors of the cosmological parameters has
the smallest eigenvalue corresponding to the largest variance.
To maximise this eigenvalue, which is to minimise the error on this
eigenvector, the shear correlation on large angular scales up to
several degree has to be measured. The dependence on the sampling of
shear correlation is different for the correlation function $\xi_{\rm
  tot}$ and the aperture mass dispersion $\langle M_{\rm ap}^2
\rangle$. For $\xi_{\rm tot}$, a survey consisting of sparse patches
with a small cosmic variance yields the smallest errors. Using
$\langle M_{\rm ap}^2 \rangle$, dense and slightly smaller patches are
preferred.

We consider various combinations of the cosmological parameters
$\Omegam, \Gamma, \sigma_8, n_{\rm s}, \Omega_\Lambda$ and the source
galaxy redshift characteristics $z_0$ and $\beta$ (see
Sect.~\ref{sec:shear_est}). If not more than three parameters are to
be determined from our survey (the other parameters being fixed), the
dominant eigenvector of the Fisher matrix comprises more than 99$\,$\%
of the parameter errors. For more than three free parameters, the
first two principal components are responsible for most of the
parameter errors.

Changing the number density of source galaxies, the depth of the
survey, the intrinsic ellipticity dispersion or the total survey area
simply causes a scaling of the eigenvalues. The optimal survey setting
is not affected by changes in those parameters.  Furthermore,
introducing external priors on some parameters does not result in a
different optimal setting. However, this optimum becomes less
pronounced so that optimisation of the survey design will be less
important. If a flat Universe is taken as a prior, the
dependence on the survey geometry is much weaker than for the
prior $\Omega_\Lambda = {\rm const}$.

The dominant principal component shows an unexpected behaviour when
the survey depth $z_0$ is varied. The error on the corresponding
parameter combination increases for increasing $z_0$ in the range of
$z_0 = 0.8 \ldots 1.2$. However, this
effect is only local and disappears when considering the likelihood
function, which shows strong non-Gaussian features not present in the
Fisher matrix.

As survey sizes increase it will be desirable to estimate an
increasing number of parameters independently from weak lensing
surveys alone such as the equation of state of dark energy or its
variation with redshift (Hu 1999; Heavens 2003).  However, parameter
near-degeneracies inherent in weak lensing observables make additional
data from different cosmology experiments very valuable in
breaking such degeneracies.  On the other hand, priors from CMB or SN
observations can be used in the future to optimise the design
of smaller surveys by using principal components of the joint Fisher
matrix (e.g.\ Crittenden \& Pogosian 2005 provide results of joint PCA
of different data sets in a similar context).

We conclude that existing tools such as generalised eigenmode analyses
(Karhunen 1947; Lo\`eve 1948; Vogeley \& Szalay 1996; Matsubara et al. 2000;
Kilbinger \& Munshi 2005) and principal component analyses which we
have studied here will be very useful in optimising future weak lensing surveys.

\section*{Acknowledgments}

DM acknowledges useful discussions with Patrick Valageas, Lindsay
King, George Efstathiou and Alan Heavens. It is a pleasure for DM to
acknowledge many fruitful discussions with members of the Cambridge
Planck Analysis Center (CPAC). MK wishes to thank Peter Schneider for
valuable discussions and comments on the manuscript. We thank
  the anonymous referee for helpful suggestions.
DM was supported
by PPARC of grant RG28936. MK was supported by the German Ministry for
Science and Education (BMBF) through the DLR under the project 50 OR
0106, and by the Deutsche Forschungsgemeinschaft under the project
SCHN 342/3--1.

\end{document}